\documentclass[aps,prd,preprintnumbers,nofootinbib,floatfix]{revtex4}
\usepackage{subcaption}
\usepackage{xcolor}
\usepackage{graphicx}
\usepackage{wrapfig}
\usepackage{amsmath}
\usepackage{amsfonts}
\usepackage{amssymb}
\usepackage{multirow}
\usepackage{slashed}
\usepackage{physics}
\usepackage{float}
\usepackage{dcolumn}
\usepackage{comment}
\usepackage[colorlinks=true, linkcolor=blue, citecolor=blue, urlcolor=blue]{hyperref}

\newcommand\befs{\begin{figure*}}
\newcommand\eefs[1]{\label{fig:#1}\end{figure*}}
\newcommand\bef{\begin{figure}}
\newcommand\eef{\end{figure}}
\newcommand\beq{\begin{equation}}
\newcommand\eeq[1]{\label{#1}\end{equation}}
\newcommand\beqa{\begin{eqnarray}}
\newcommand\eeqa[1]{\label{#1}\end{eqnarray}}
\newcommand\bet{\begin{table}}
\newcommand\eet{\end{table}}
\newcommand\bets{\begin{table*}}
\newcommand\eets[1]{\label{tb:#1}\end{table*}}
\newcommand{\be}{\begin{equation}}
\newcommand{\ee}{\end{equation}}
\newcommand{\bea}{\begin{eqnarray}}
\newcommand{\eea}{\end{eqnarray}}

\newcommand\fgn[1]{Fig.\ \ref{fg:#1}}
\newcommand\eqn[1]{Eq.\ (\ref{#1})}
\newcommand\scn[1]{Section \ref{sec:#1}}

\newcommand\tbn[1]{Table \ref{tb:#1}}

\begin{document}

\date{\today}

\title{Anomalous behavior of Wilson fermions in the presence of monopoles }

\author{Manuel\ \surname{Cortina}}
\email{mcort086@fiu.edu}
\author{Rajamani\ \surname{Narayanan}}
\email{rajamani.narayanan@fiu.edu}
\author{Ray\ \surname{Romero}}
\email{rrome071@fiu.edu}
\affiliation{Department of Physics, Florida International University, Miami, FL 33199}

\begin{abstract}
Three dimensional compact $U(1)$ gauge theory supports a finite density of monopoles at a fixed lattice gauge coupling. After showing evidence for the expected dependence on the lattice coupling of the density of monopoles, we study the low lying spectrum of a two flavors of two-component Wilson fermions as a function of the bare Wilson mass. The Wilson-Dirac operator that realizes the two flavors are hermitian conjugate of each other to restore parity invariance and the commutativity of the two operators expected in the typical continuum limit is not the case in the presence of monopoles for all values of the lattice gauge coupling. This results in an anomalous behavior where we see clear evidence for the presence of very small eigenvalues for a wide range of a particular sign of the bare Wilson mass. These eigenmodes are typically in one-to-one correspondence with the number of monopoles as defined by the gauge field and the spectrum associated with these modes are separated by a gap from the bulk in the asymptotic regime of the lattice gauge coupling. The eigenvalues come in pairs around zero and the eigenvalues of one sign couple to the monopoles while the modes of the opposite sign couple to the anti-monopole implying flavor symmetry breaking.
\end{abstract}

\maketitle

\section{Introduction and summary}

Wilson gauge action allows the presence of monopoles~\cite{Polyakov:1975rs,Polyakov:1976fu} due to the compact nature of the gauge fields.
These are fundamentally singular configurations and are expected to strongly affect the continuum limit. The Villain gauge action also allows monopoles and is easier to study analytically~\cite{Gopfert:1981er} since it can be mapped to a scalar field theory with the scalar field restricted to take on integer values. This allows for the computation of the mass gap, $m_D$,~\cite{Savit:1977fw,Banks:1977cc,Ukawa:1979yv,Muller:1981ia} given by
\be
M = am_D = \sqrt{8\pi^2\beta} e^{-c\beta};\qquad c = 0.2527\pi^2=2.494,\label{massgap}
\ee
where $a$ is the lattice spacing and $\beta$ is the lattice gauge coupling appearing in the Wilson gauge action to be defined in \scn{prelim}.
The analysis of the string tension, $\alpha$,
in~\cite{Gopfert:1981er} results in a lower bound that can be approximated to an equality in the semi-classical approximation,
\be
\Sigma=a^2\alpha =  \frac{8} {\sqrt{2\pi^2 \beta}}e^{-c\beta}.\label{sigma}
\ee
Assuming the monopoles form a random dilute gas~\cite{Athenodorou:2018sab}, the typical distance between monopoles, $L_M$, obeys $L^3_M \sim L_\Sigma^2 L_M = \frac{1}{\Sigma M}$. Therefore, the density of monopoles is
\be
\frac{1}{L_M^3} \sim \Sigma M  \sim 16\ e^{-2 c\beta}.\label{mden}
\ee

A numerical analysis of the integer valued scalar field theory was performed very early~\cite{Karliner:1983ab} and a reasonable agreement was found with~\eqn{massgap} resulting in an estimate for $c$. This estimate cannot be expected to be as precise as the value obtained in~\cite{Muller:1981ia} since that is an outcome of a lattice momentum integral. A recent numerical computation~\cite{Athenodorou:2018sab} of the string tension using the Wilson gauge action and correlation functions of Polyakov loops that wind around the torus provide a precise estimate of $c$ obtained from~\eqn{sigma}.

Monopoles can be measured for any compact gauge action using the map between the compact plaquette variable and an angle in the range $(-\pi,\pi]$~\cite{DeGrand:1980eq}. A natural definition arises from the character expansion that defines the Villain action and the two methods were compared in~\cite{Schram:1993gq} and found to be in good agreement. Furthermore, a behavior of the monopole density consistent with~\eqn{mden} was also found but a value of the constant, $c$, was not extracted. An extensive analysis of the behavior of monopoles using the Villain action can be found in~\cite{Chernodub:2001ws} including a study of the density of monopoles as a function of the lattice coupling.

The possibility of different continuum limits for the lattice gauge theory with the Wilson gauge action and the associated physical consequences has been beautifully discussed in~\cite{Athenodorou:2018sab}. We simply recast this discussion here by considering the continuum theory on a finite periodic box, $\ell^3$, which is kept fixed in the approach to the continuum limit
and extracting $\beta$ as a function of $a$ while keeping a certain physical quantity fixed.  In order to define the lattice spacing, we say that $\ell = aL$. 
Then \eqn{massgap}, \eqn{sigma} and \eqn{mden} (converted to an equation for the total number of monopoles) can be rewritten as
\be
m\ell = ML =  \sqrt{8\pi^2 \beta} L e^{- c \beta};\qquad \alpha \ell^2 = \Sigma L^2  
=\frac{8}{\sqrt{2\pi^2\beta}} L^2 e^{- c \beta};\qquad \frac{\ell^3}{\ell_M^3}=\frac{L^3}{L_M^3} = 16 L^3 e^{-2 c\beta}.
\ee
The right-hand side of each of these equations tells us how we should change $\beta$ as a function of $L$ to keep the left-hand side constant. These result in
\be
L = \frac{m\ell}{\sqrt{8\pi^2\beta}}e^{ c\beta};\quad {\rm or}\quad L^2 = \frac{\alpha\ell^2\sqrt{2\pi^2\beta}}{8}e^{c\beta};
\quad {\rm or}\quad L^3 = \frac{\ell^3}{16 \ell_M^3}e^{2c\beta};\qquad L\to\infty\quad\Rightarrow\quad \beta\to\infty.
\ee
If we wish to keep the mass gap, $m\ell$, fixed, then we see that 
\be
\left.\lim_{L\to\infty}\right|_{m\ell} \alpha\ell^2 = \left.\lim_{L\to\infty}\right|_{m\ell} \frac{\ell^3}{\ell_M^3}=\infty.
\ee
If we wish to keep the string tension, $\alpha\ell^2$, fixed, then we see that 
\be
\left.\lim_{L\to\infty}\right|_{\alpha\ell^2} m\ell = \left.\lim_{L\to\infty}\right|_{m\ell} \frac{\ell^3}{\ell_M^3}= 0.
\ee
If we wish to keep the number of monopoles, $\frac{\ell^3}{\ell_M^3}$, fixed, we have
\be
\left.\lim_{\beta\to\infty}\right|_{\frac{\ell^3}{\ell_M^3}} m\ell=0;\qquad 
\left.\lim_{\beta\to\infty} \right|_{\frac{\ell^3}{\ell_M^3}} \alpha^2\ell=\infty.
\ee
The above continuum limits should be contrasted with, $\beta=\frac{2L}{\ell}$, that arises from dimensional analysis and should be valid if the theory is super-renormalizable wherein
\be
\lim_{L\to\infty} m\ell = \lim_{L\to\infty} \alpha\ell^2 = \lim_{L\to\infty} \frac{\ell^2}{\ell_M^3} = 0,\label{convcont}
\ee
at any finite physical volume.

As discussed in~\cite{Athenodorou:2018sab}, we can couple a physical charged particle only in the limit where there is no mass gap, the potential is Coulomb like and no monopoles are present (c.f.~\eqn{convcont}). This limit should be the same as the theory with a non-compact lattice gauge action and the fermions couple to the compact parallel transporter on the link defined via the non-compact gauge field. The charge of the fermion is only relevant if we have more than one flavor and in such a case all relative charges have to be integers for a proper mapping from the non-compact link variable to the compact parallel transporter. Initial studies using non-compact gauge action can be found in~\cite{Hands:1989mv,Hands:2002dv,Hands:2004bh} and recent studies of parity invariant theories with even number of two component fermion flavors with Wilson fermions and overlap fermions have resulted in several physics results. The theories are scale invariant for all possible number of flavors~\cite{Karthik:2015sgq,Karthik:2016ppr} and there is evidence for the extended $O(4)$ symmetry in the case of two flavors~\cite{Karthik:2017hol}. It is possible to insert a single monopole-anti-monopole pair within the non-compact gauge action and the scaling dimensions of the monopole operator with different fluxes have been computed~\cite{Karthik:2018rcg,Karthik:2024ffr}.

As mentioned before, parity can be preserved on the lattice for both Wilson fermions and overlap fermions if the gauge action is non-compact and we have an even number of two-component fermion flavors. The operators for a pair of fermions that cancel the parity anomaly commute in the case of overlap fermions but  not for Wilson fermions. Naively, the commutator should vanish in the continuum for smooth fields but this is not the case for monopoles.  An analysis of the fermion spectrum in a fixed gauge fields background of a well separated monopole and anti-monopole gave some unexpected results~\cite{Karthik:2019jds}. A parity invariant pair of Wilson fermions has a spectrum that is not symmetric about the massless point. A pair of anomalously small eigenvalues that goes to zero on an infinite lattice appears for a wide range of one sign of the mass. Since the Wilson-Dirac operator is the kernel for the overlap-Dirac operator, it does affect the spectrum of a parity invariant pair of overlap fermions in spite of the fact the pair of overlap fermion operators commute. The Wilson mass parameter that plays the role of a regulator has an effect on the numerical values of the spectrum and it also raises the question in principle of the infinite lattice limit.

Our aim in this paper is to study the connection between the monopoles and Wilson fermion spectrum using the compact Wilson gauge action. Fermions themselves do not play a dynamical role. We will start with a study of the properties of monopoles by measuring the monopoles using the relation between the compact plaquette and the associated angle~\cite{DeGrand:1980eq}. We will find results that are consistent with previous results~\cite{Schram:1993gq,Chernodub:2001ws} and our precision will be good enough to show agreement with~\eqn{mden}. Our estimate will be $c=2.49(4)$. The coefficient in front of the exponent in \eqn{mden} uses two approximations. It says there is no additional factor in relating the monopole density to string tension and the mass gap and the string tension itself has a coeffcient in \eqn{sigma} that comes out of a semi-classical approximation. Numerical computation of $\Sigma\sqrt{\beta}$ in~\cite{Athenodorou:2018sab} gives a value of $11.5(6)$ which is much higher that $\frac{8}{\sqrt{2\pi^2}}$. Our estimate for the coefficient in front of the exponent in \eqn{mden} will be $35(6)$. We will then study the dependence of the action density, $\rho_S(\beta)$, as a function of the monopole density, $\rho_M(\beta)$, and find that 
$\rho_S(\beta) = 4.01(3)\rho_M(\beta)$ as $\beta\to\infty$.  To further lend support to a random gas of monopoles, we will compare the number of pairs where each entity in the pair can be a monopole (positive charge) or an anti-monopole (negative charge) to the number of pairs if it were a random gas. We will find a repulsion between like charges and attraction between unlike charges at short distances that remains fixed as the lattice size goes to infinity at a fixed lattice coupling. Since the density of monopoles decreases as the lattice coupling increases, the range of distances at which we see appreciable attraction or repulsion will increase with the lattice coupling. Since the mean distance between charges will grow with lattice size at a fixed lattice spacing due to  a random dilute gas like behavior, the effect of attraction and repulsion is a microscopic effect. With this place, we will move on to the analysis of the spectral properties of a pair of Wilson fermions with a mass parameter that is equal and opposite. We will find that the results depend on the sign of the fermion mass and we will show the following:
\begin{itemize}
\item The spectrum will be symmetric about zero at every value of the Wilson mass parameter and this is by construction.
\item The spectrum will have anomalously small eigenvalues when the mass parameter switches sign. The number of such small eigenvalues will be essentially in one-to-one correspondence with the number of monopoles based  on a function of the gauge field background.
\item We will show that the eigenvectors are localized at the location of the monopole or anti-monopole when the background is a well separated monopole and anti-monopole.
\item The local property of the eigenvectors associated with the anomalously small eigenvalues will result in eigenvectors that are localized at a single monopole or anti-monopole location.
\item The anomalously small eigenvalues will be well separated from the rest of the spectrum for the sign of mass where they are present. Furthermore, these eigenvalues will approach zero exponentially as the size of the lattice is increased at a fixed lattice coupling and Wilson mass. The rate of this exponential fall-off will depend linearly on the Wilson mass parameter and will approach close to zero as the Wilson mass parameter goes to zero. The slope associated with this linear dependence will increase with the lattice gauge coupling.
\end{itemize}

A finite density of monopoles produces a finite density of zero eigenvalues for a pair of Wilson-Dirac operators in an infinite lattice at all values of a particular sign of Wilson mass. This results in an infinite fermion bilinear condensate for a range of the Wilson mass parameter. This is an effect due to the singular nature of the monopoles and the physical effect will be different for Wilson and overlap fermions. We do not perform any computation with overlap fermions in this paper but the results from~\cite{Karthik:2019jds} suggest that the small eigenvalues of the overlap-Dirac operator only present at the massless point will not result in a condensate. We will provide the best supporting evidence for this using Wilson fermions by approaching the massless point. We start with the necessary technical details in \scn{prelim}. The results for the monopoles are presented in \scn{monopoles} and the results for the  Wilson fermions are presented in \scn{wilson}. These are followed by a concluding section.

\section{Technical details of the lattice formalism}\label{sec:prelim}

We will work on a $L^3$ lattice and we will assume $L$ is even.
The Euclidean partition function is
\be
Z = \int [dU_\mu({\bf n})] e^{-\beta S_g},
\ee
where $S_g$ is the action and $\beta$ is the lattice gauge coupling. A site on the lattice is ${\bf n}=(n_1,n_2,n_3)$ with $0\le n_i< L$ and
the gauge link  connecting ${\bf n}$ and ${\bf n}+\hat\mu$; $\mu=1,2,3$ is
\be
U_\mu({\bf n}) = e^{i\theta_\mu({\bf n})}.
\ee
The gauge invariant plaquette variable is
\be
P_{\mu\nu}({\bf n}) \equiv e^{i\theta_{\mu\nu}({\bf n})} =
U_\mu({\bf n})U_\nu({\bf n}+\hat\mu)U^*_\mu({\bf n}+\hat\nu)U^*_\nu({\bf n});\qquad P_{\nu\mu}({\bf n}) = P^*_{\mu\nu}({\bf n}).\label{cplaq}
\ee
It will be useful to define real quantities,
\bea
s_{\mu\nu}({\bf n}) &=& -s_{\nu\mu}({\bf n}) = \frac{1}{2i} \left [P_{\mu\nu}({\bf n}) - P_{\nu\mu}({\bf n})\right] = \sin\theta_{\mu\nu}({\bf n});\cr
c_{\mu\nu}({\bf n}) &=& c_{\nu\mu}({\bf n}) = \frac{1}{2} \left [P_{\mu\nu}({\bf n}) + P_{\nu\mu}({\bf n})\right]= \cos\theta_{\mu\nu}({\bf n}).\label{csplaq}
\eea
The compact gauge action is the Wilson action,
\be
S_g=\sum_{\bf n}\sum_{\mu<\nu}\left[1-c_{\mu\nu}({\bf n}) \right].\label{waction}
\ee

A discrete HYP smearing~\cite{Hasenfratz:2001hp,Hasenfratz:2007rf} step is implemented by
\be
U_\mu^{(j+1)}({\bf n}) = U_\mu^{(j)}({\bf n}) X_\mu^{(j)}({\bf n});\qquad X_\mu^{(j)}({\bf n})=e^{iQ_\mu^{(j)}({\bf n})}; \qquad
Q_\mu^{(j)}({\bf n}) =   -\frac{s_1}{4}  \sum_{\nu\ne\mu} \bar\Delta_\nu s^{(j)}_{\mu\nu}({\bf n}).
\label{smear}
\ee
Observables will depend on smeared links to reduce finite lattice spacing effects.
\subsection{Parity}
Parity will be relevant for us and we provide the details of this transformation on a finite periodic lattice.
The parity operator is defined
as
\be
(P\psi)({\bf n}) = \psi^p({\bf n}) = \psi({\bf L} - {\bf n});\qquad {\bf L}=(L,L,L)
;\qquad
P^\dagger = P;\qquad P^2=I.
\ee

Let us define the parallel transporter (an unitary operator) in the $\mu$ direction by its action on a scalar field as
\be
(T_\mu(U) \psi)({\bf n}) = U_\mu({\bf n})\psi ({\bf n}+\hat\mu).
\ee
Then 
\be
PT_\mu(U)P = T^\dagger_\mu(U^p)\quad\Rightarrow\quad  U^p_\mu({\bf n})=e^{i\theta^p({\bf n})};\qquad  \theta^p_\mu({\bf n})=-\theta_\mu({\bf L}-{\bf n}-\hat\mu).\label{parityt}
\ee
and 
\be
P_{\mu\nu}^p({\bf n}) = P_{\mu\nu}({\bf L} - {\bf n}-\hat\mu-\hat\nu)
\label{plaqpar}
\ee
showing that the action of a gauge configuration and its parity partner have the same action. In addition, the parity relation applies to the smeared links.

\subsection{Fermions}
We will use the Sheikhoslami-Wohlert~\cite{Sheikholeslami:1985ij} improved Wilson-Dirac operator to study the theory with dynamical fermions. In addition, as mentioned before, the operator will depend on the smeared links but we will just use $U$ to define the operator.
\subsubsection{Na\"ive Dirac operator}
Let $\psi_\alpha(n_1,n_2,n_3)$ with $\alpha=1,2$ be the fermionic field where $\alpha$ denotes the spin index.
The Hermitian Pauli matrices are
\be
\sigma_1 = \begin{pmatrix} 0 & 1 \cr 1 & 0 \end{pmatrix};\quad
\sigma_2 = \begin{pmatrix} 0 & -i \cr i & 0 \end{pmatrix};\quad
\sigma_3 = \begin{pmatrix} 1 & 0 \cr 0 & -1 \end{pmatrix};\qquad \sigma_j = i \epsilon_{jkl}\sigma_k \sigma_l;\qquad \sigma_j^2=1;\ j=1,2,3.
\ee

The na\"ive Dirac operator is defined by
\be
{D}(U) = \frac{1}{2}\sum_{\mu=1}^3\sigma_\mu \left(T_\mu(U) - T_\mu^\dagger(U)\right);\qquad {D}^\dagger(U)  = -{D} (U).
\ee
The eigenvalues of ${D}(U)$ are purely imaginary. 

Under parity, we can use \eqn{parityt} to find
\be
P{D}(U)P = -{D}(U^p).\label{parityn}
\ee

\subsubsection{Improved Wilson-Dirac operator}
The Wilson mass term is given by
\be
B(U) = 3+m_w -\frac{1}{2}\sum_{\mu=1}^3\left(T_\mu(U) + T^\dagger_\mu(U)\right);\qquad B^\dagger(U) = B(U),
\ee
and $m_w$ is the Wilson mass parameter.
Under parity, we can use \eqn{parityt} to find
\be
PB(U)P = B(U^p).\label{parityb}
\ee
The Wilson-Dirac operator is given by
\be
{D}_w(U) = {D}(U) + B(U).
\ee
Then
\be
{D}^\dagger_w(U) = -{D}(U) + B(U).
\ee
Therefore, the spectrum of ${D}_w(U)$ is complex. 

To improve the Wilson-Dirac operator, we add
 the Sheikholeslami-Wohlert operator, $C_{\rm sw}(U)$. Its action on a fermion field is given by
\be
\left[C_{\rm sw}(U)\psi\right]({\bf n}) = i\frac{c_{\rm sw}}{8} \sum_{\mu\nu\rho} \epsilon_{\mu\nu\rho} C_{\mu\nu}({\bf n}) \sigma_\rho\psi({\bf n});\quad
C_{\mu\nu}({\bf n}) = P_{\mu\nu}({\bf n}) +P_{\mu\nu}({\bf n}-\hat\mu)+ P_{\mu\nu}({\bf n}-\hat\nu)+ P_{\mu\nu}({\bf n}-\hat\mu-\hat\nu).
\ee
Note that
\be
C^*_{\mu\nu}({\bf  n}) = C_{\nu\mu}({\bf n})\quad\Rightarrow\quad C^\dagger_{\rm sw}(U) = C_{\rm sw}(U).
\ee
Using \eqn{plaqpar}, we see that
\be
PC_{\rm sw}(U)P = C_{\rm sw}(U^p).
\ee
The improved Wilson-Dirac-Sheikholeslami-Wohlert operator is
\be
D_{i}(U) = D_w(U) + C_{\rm sw}(U) = D(U)+B(U)+C_{\rm sw}(U).
\ee
Combining this with \eqn{parityn} and \eqn{parityb}, we have
\be
P{D}_i(U)P = {D}^\dagger_i(U^p).
\ee
As such, we cannot relate the spectrum of ${D}_i(U)$ and ${D}_i(U^p)$ but we can say that
\be
\det [{D}_i(U)] = {\det}^*[{D}_i(U^p)].
\ee
Therefore Wilson fermions sees a parity anomaly~\cite{Coste:1989wf}. We therefore consider a pair of fermion flavors and write the fermion operator on the pair of fermions as
\be
{W}(U) = \begin{pmatrix} 0 & {D}_i(U) \cr -{D}^\dagger_i(U)& 0 \end{pmatrix}
\quad\Rightarrow\quad {W}(U^p) = \begin{pmatrix} 0 & P{D}^\dagger_i(U) P \cr -P{D}_i(U)P & 0 \end{pmatrix}.\label{woper}
\ee
The fermion determinant for the pair of fermions is parity invariant. The spectrum of $W$ is symmetric about zero but the spectrum at $m_w$ is not simply related to the one at $-m_w$. If we can assume that the gauge field dependent part of $B(U)$  and $C_{\rm sw}(U)$ are irrelevant compared to $D(U)$ in the continuum limit we can think of the pair formed by $D_i(U)$ and $D^\dagger_i(U)$ to have equal and opposite masses and we can expect the spectrum in the continuum limit to be symmetric around $m_w=0$. Our results will show that the spectrum has anomalously small eigenvalues separated by a gap from the rest of the spectrum when monopoles are present in the configuration and $m_w < 0$. Furthermore, the number of such small eigenvalues can be used as a measure of the total number of monopoles. 
We note that the Wilson kernel that appears in the overlap-Dirac operator~\cite{Karthik:2016ppr} has $-2 < m_w < 0$.

\subsection{Observables}

We use the standard heat-bath and over-relaxation algorithms to perform updates on our gauge fields. One update constitutes of 1 heat-bath step followed by 4 over-relaxation steps. The number of updates between measurements depend on $\beta$. The low lying spectrum of $W(U)$ was computed using the Arnoldi algorithm available via ARPACK~\cite{arpack} and sufficient number of eigenvalues were computed to include all anomalously small eigenvalues. \tbn{Simtab} provides the list of $\beta$ and $L$ that were used to generate gauge fields. Due to the computational intensity of the eigenvalues, these were performed only on a subset of $(\beta,L)$ as shown in \tbn{Simtab}.  Only even values of $L$ were used. Fermions obey periodic boundary conditions in all three directions. All observables were measured with $50$ HYP smearing steps and the smearing parameter in \eqn{smear} was set to $s_1=0.05$.

\begin{table}[htbp]
\centering
\caption{Simulation parameters for each $\beta$}
\label{tb:Simtab}
\begin{tabular}{c c c c}
\hline
$\beta$ & $L$ range for  & $L$ range for measurements of eigenvalues  & Update steps \\
 &  gauge field measurements & of the Wilson-Dirac operator &  \\
\hline
1.500 & 16--46 & & 40 \\
1.625 & 16--48 & 16--32 & 40 \\
1.750 & 16--46 & 16--40 & 40 \\
1.875 & 16--48 & 16--38 & 40 \\
2.000 & 16--72 & 16--38 & 40 \\
2.125 & 16--72 & 16--38 & 40 \\
2.250 & 16--72 & 16--38 & 40 \\
2.375 & 16--72 & 16--38 & 40 \\
2.500 & 16--72 & 16--38 & 60 \\
2.625 & 16--72 & 16--38 & 60 \\
2.750 & 16--72 & 32--44 & 60 \\
2.875 & 16--72 & & 60 \\
3.000 & 16--72 & & 80 \\
\hline
\end{tabular}
\end{table}

\bef 
\includegraphics[scale=0.45]{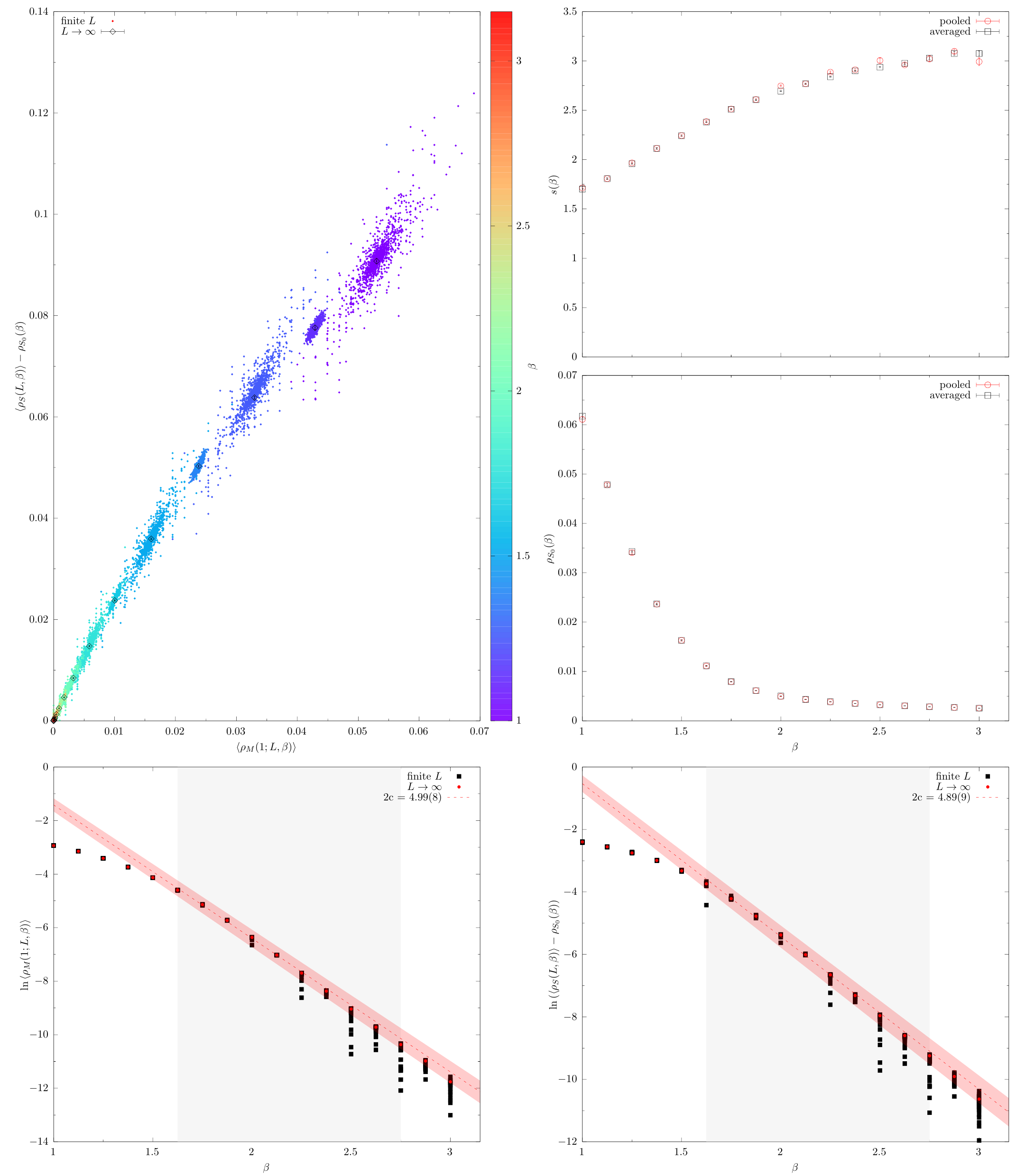}
\caption{Behavior of the monopole density and associated action density as a function of the lattice gauge coupling, $\beta$.} 
\label{fg:action_mono_den_1}
\eef

\subsubsection{Monopoles}\label{sec:monoobs}
We define~\cite{DeGrand:1980eq} the monopole charge at a particular site by
\be
2\pi M({\bf n}) = \sum_{f} \theta_f({\bf n});\qquad -\pi<\theta_f({\bf n}) \le \pi\label{gmono}
\ee
where $f$ is the sum over the six faces of cube located at ${\bf n}$ and the three axes being in the three positive direction and
$ e^{i\theta_f({\bf n})}$ is the plaquette value using the smeared fields on each of the six faces such that the orientation of each plaquette points out of the cube.
We will find $M({\bf n})$ to be an integer in $[-3,3]$. Let $N(p)$ with $p\in [-3,3]$ be the number of cubes that resulted in $M({\bf n})=p$ on a $L^3$ lattice at a fixed $\beta$. We define
\be
\rho_M(p;L,\beta) = \frac{N(p;L,\beta)}{L^3},
\ee
as the density of monopoles in a fixed gauge field background at a given $\beta$ and $L$. Let us define the action density as
\be
\rho_S(L,\beta) = \frac{S_g}{3L^3}
\ee
on the same fixed gauge field background and $S_g$ is the action given in~\eqn{waction}.

Let ${\bf n}_i$ and ${\bf m}_i$; $i=1,\cdots,K$ be the locations of the positively and negatively charged unit monopoles in a single configuration. 
Let $d^+_{ij}$ and $d^-_{ij}$ be the shortest distance on the lattice between monopoles, ${\bf n}_i,{\bf n}_j$, and   anti-monopoles, ${\bf m}_i,{\bf m}_j$, respectively with $i\ne j$. Let $d^{+-}_{ij}$  be the shortest distance between monopole and anti-monopole at ${\bf n}_i,{\bf m}_j$. 
Then we define,
\be
d^\pm=\frac{1}{K(K-1)} \sum_{i\ne j} d^\pm_{ij};\qquad d^{+-}=\frac{1}{K^2} \sum_{i, j} d^{+-}_{ij}.\label{meandis}
\ee
$d^\pm$ are defined only if $K>1$.
We define the average distance between monopoles, anti-monopoles and between a monopole-anti-monopole pair by taking the average over an ensemble.
Let $f(r)$ be the fraction of number of pairs at a distance $r$ in an ensemble at a fixed $\beta$ and $L$. If we assume that the monopoles can be described by a dilute random gas, we can obtain a value for $f_d(r)$ which will depend on $L$ but not on the monopole density itself as long as $L$ is large enough. 
In order to compare $f(r)$ with $f_d(r)$, we define $g(r) = \frac{f(r)}{f_d(r)}$. If $g(r)$ were larger than unity we would say we find more pairs at that separation than if the pairs were randomly distributed indicating an attraction and if it were smaller than unity then we would say we find fewer pairs at that separation then as if it were randomly distributed indicating a repulsion. 
In addition, we can also compare $d^\pm$ and $d^{+-}$ to that obtained from a dilute random gas.

\bef 
\includegraphics[scale=0.45]{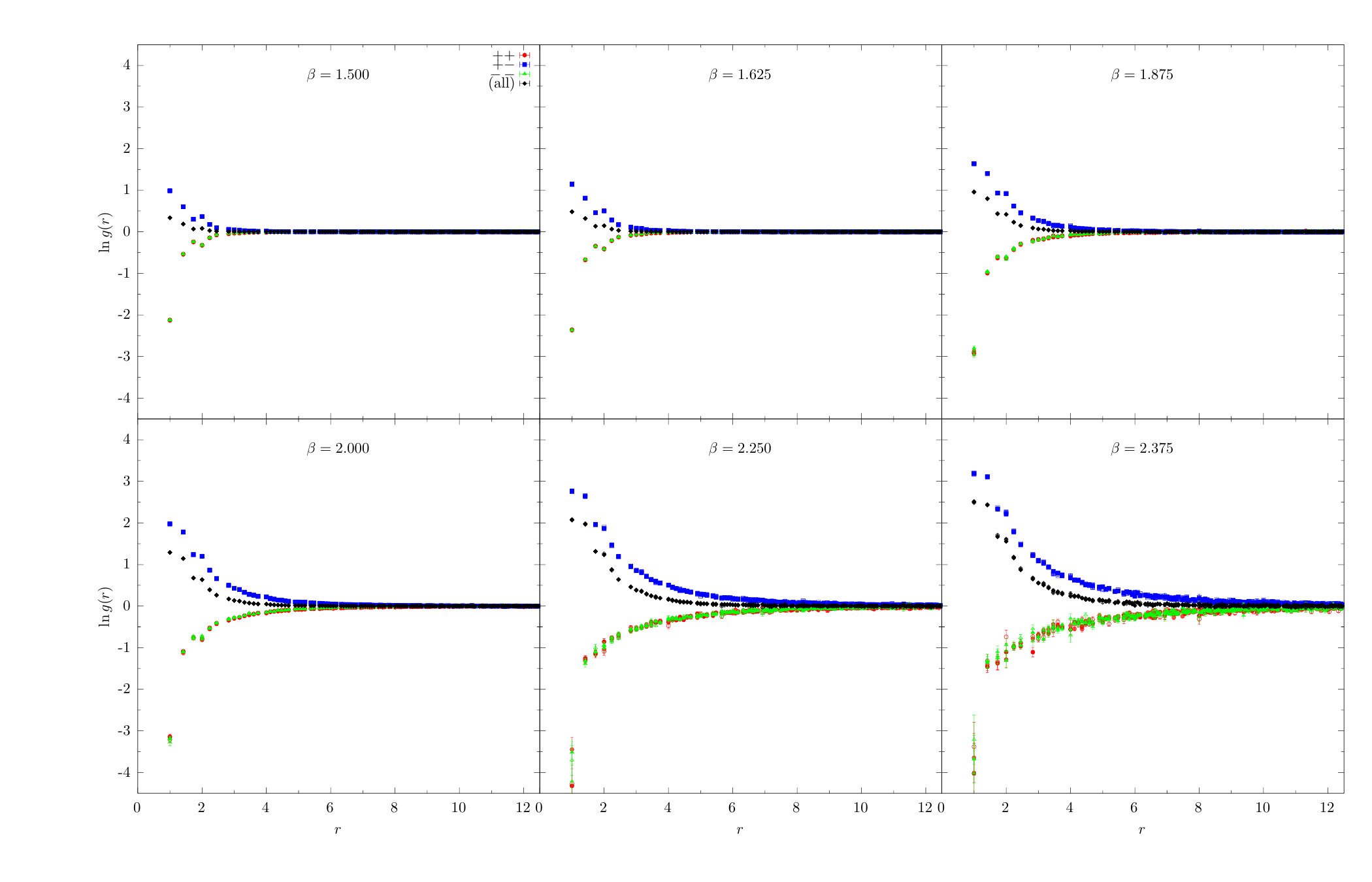}
\caption{log of g(r), ratio of pairs found at distance r to pairs at distance r if randomly distributed as a function of r.} 
\label{fg:ln_g_r}
\eef

\subsubsection{Eigenvalues of fermionic operator}
The spectral density of the Wilson-Dirac operator in a fixed gauge field background is a gauge invariant quantity and the behavior near zero eigenvalue can be useful in understanding the physics of monopoles. The eigenvalue equation (dropping the dependence on $U$ ) is
\be
W\begin{pmatrix} u_k \cr d_k \cr \end{pmatrix} = i\Lambda_k \begin{pmatrix} u_k \cr d_k \cr \end{pmatrix} \quad\Rightarrow\quad
D_iD_i^\dagger u_k = \Lambda_k^2 u_k;\quad D_i^\dagger D_i d_k = \Lambda_k^2 d_k;\quad\Rightarrow\quad
W\begin{pmatrix} \frac{D_i d_k}{\pm i\Lambda_k} \cr d_k \cr \end{pmatrix} = \pm i\Lambda_k \begin{pmatrix} \frac{D_i d_k}{\pm i\Lambda_k} \cr d_k \cr \end{pmatrix};\quad \Lambda_k > 0.\label{eigw}
\ee
Our aim is study the spectrum close to zero. The spectrum will depend on $\beta,L$ and $m_w$.

Let us show the explicit dependence on $m_w$ by writing $W(m_w)$. It will be useful for us to relate the spectrum at $m_w$ and $-(6+m_w)$~\cite{Edwards:1998gk} when $c_{\rm sw}=0$. To see this relation consider the operator
\be
(O\psi)({\bf n}) = \begin{cases} \psi({\bf n}) & \mod\left(\sum_\mu n_\mu,2\right)=0 \cr -\psi({\bf n}) & \mod\left(\sum_\mu n_\mu,2\right)\ne 0 \cr\end{cases};\qquad O^2=\mathbf{I}.
\ee
It follows that
\be
OT_\mu(U)O = -T_\mu(U);\quad OT^\dagger_\mu(U) O = -T^\dagger_\mu(U) \quad\Rightarrow\quad OD(U)O=-D(U);\quad OB(m_w,U)O = -B(-6-m_w,U)
\ee
and we have made the dependence of $m_w$ on $B$ explicit.
Therefore
\be
OW(m_w)O = W(-6-m_w)
\ee
and we see that the spectrum will be symmetric about $m_w=-3$. In addition, the spectrum at $m_w=-3$ will be doubly degenerate.

\bef
\centering
\includegraphics[scale=0.35]{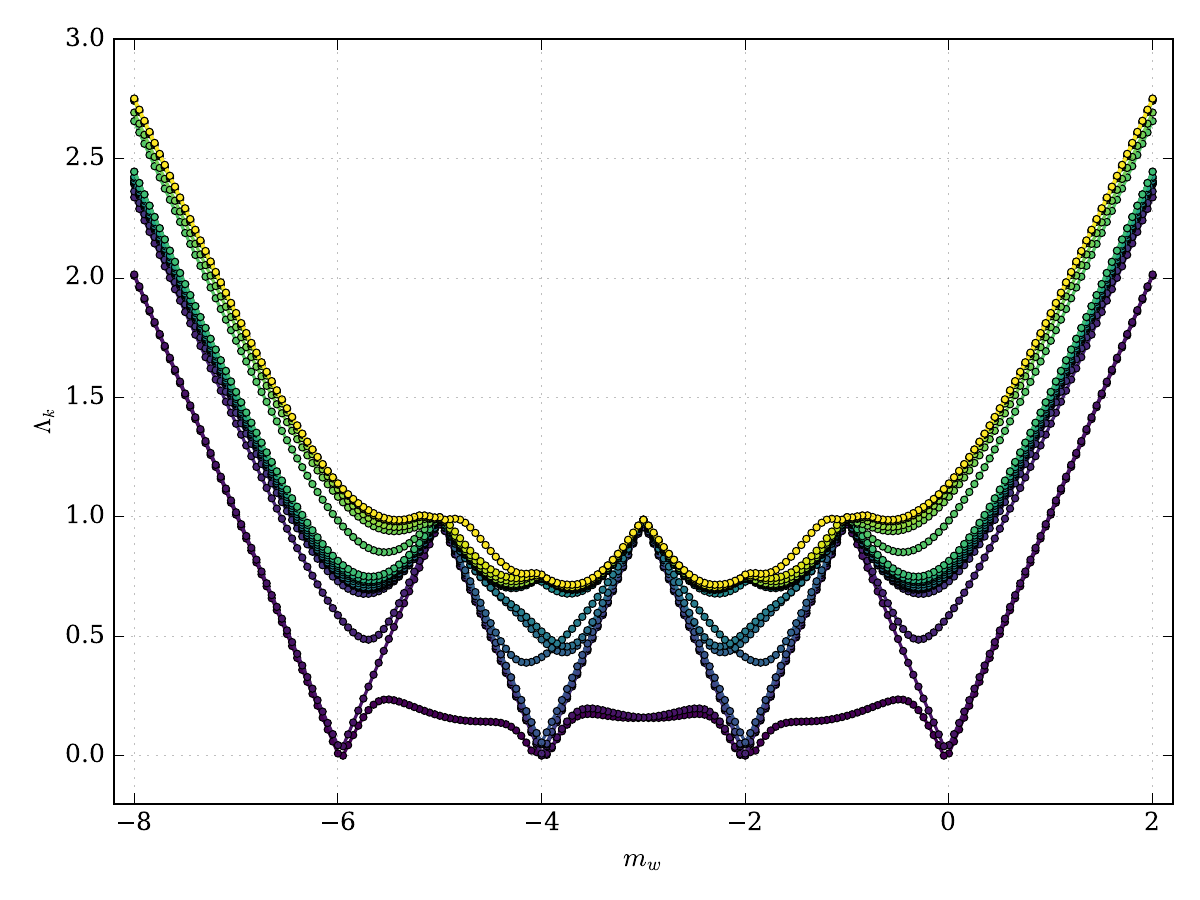}
\includegraphics[scale=0.35]{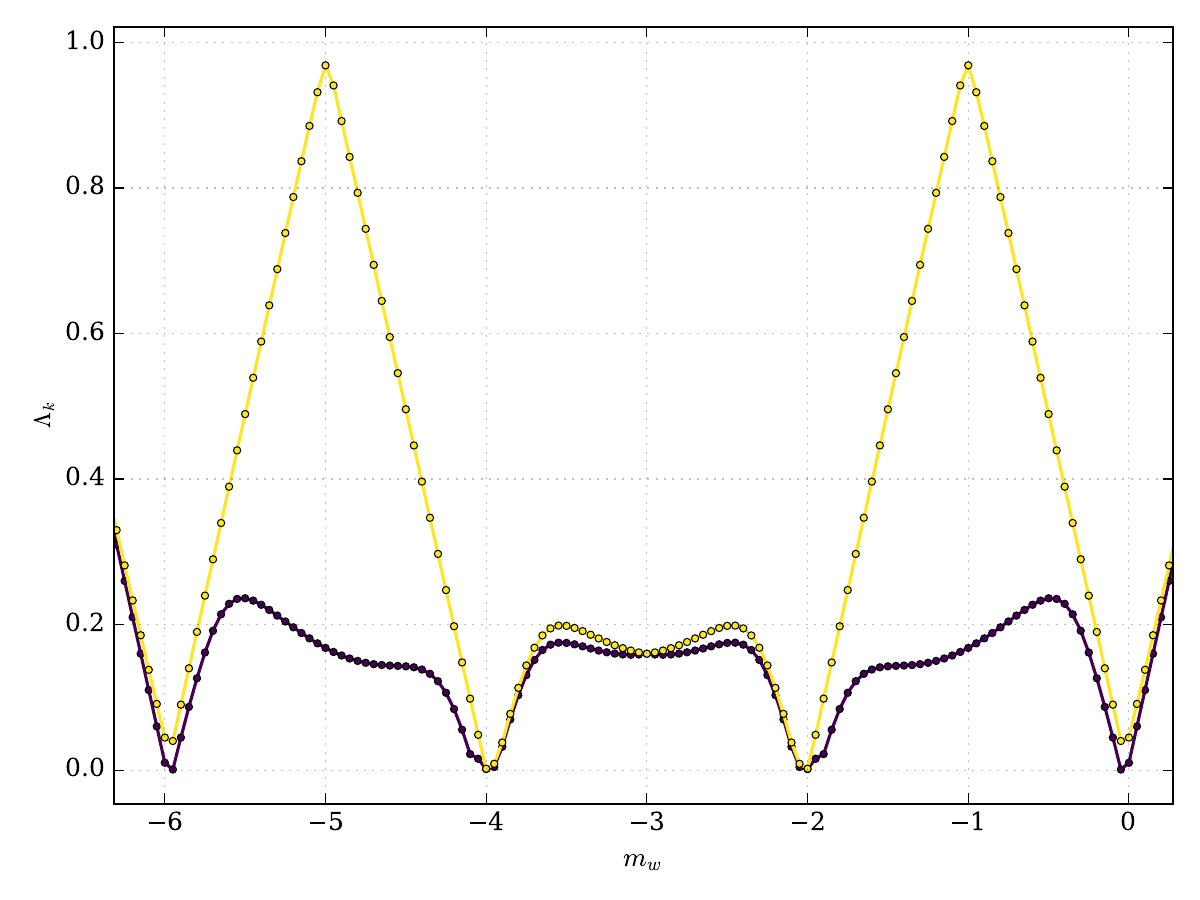}
\includegraphics[scale=0.45]{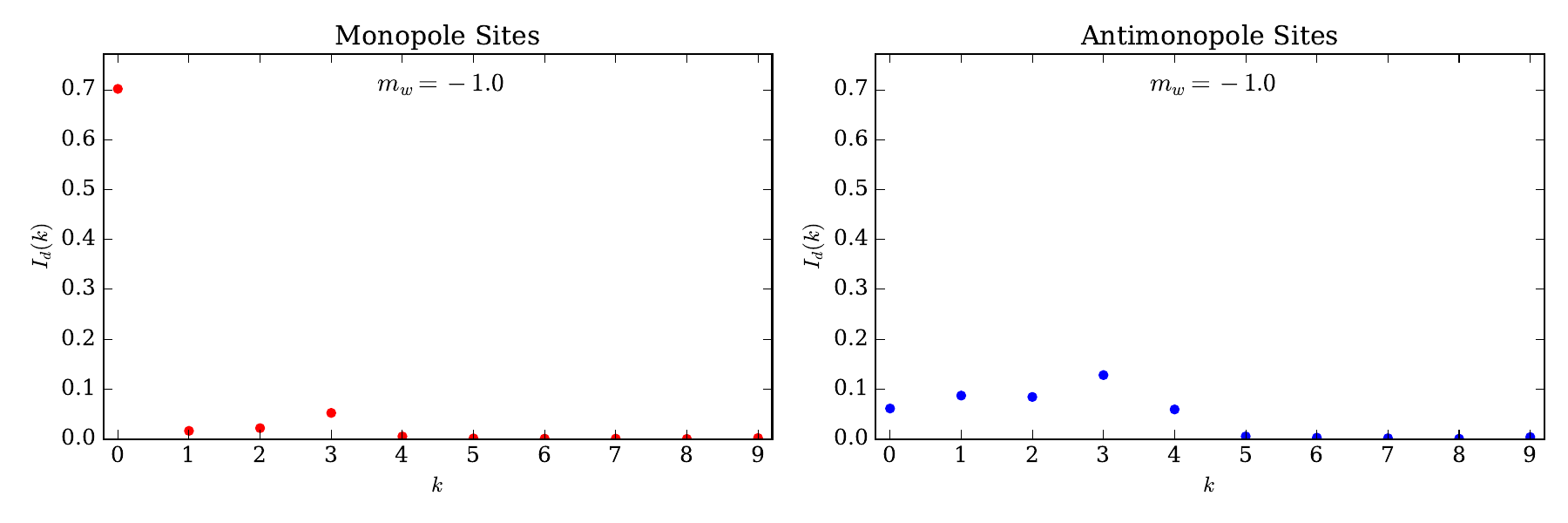}
\includegraphics[scale=0.45]{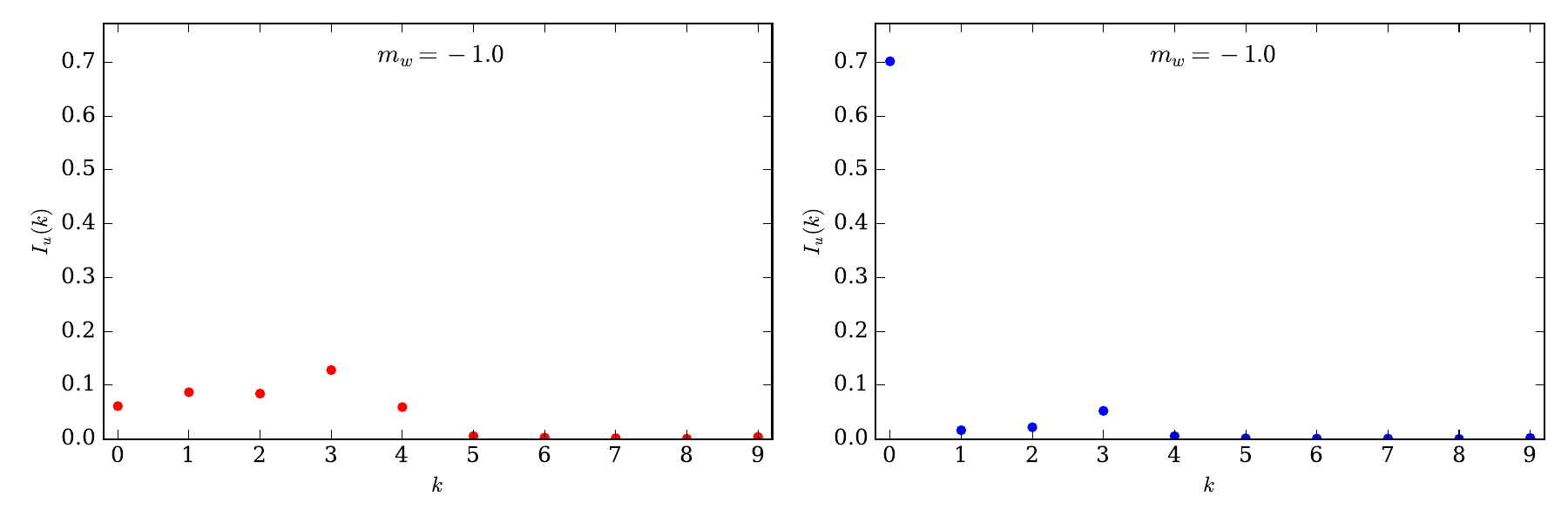}
\includegraphics[scale=0.45]{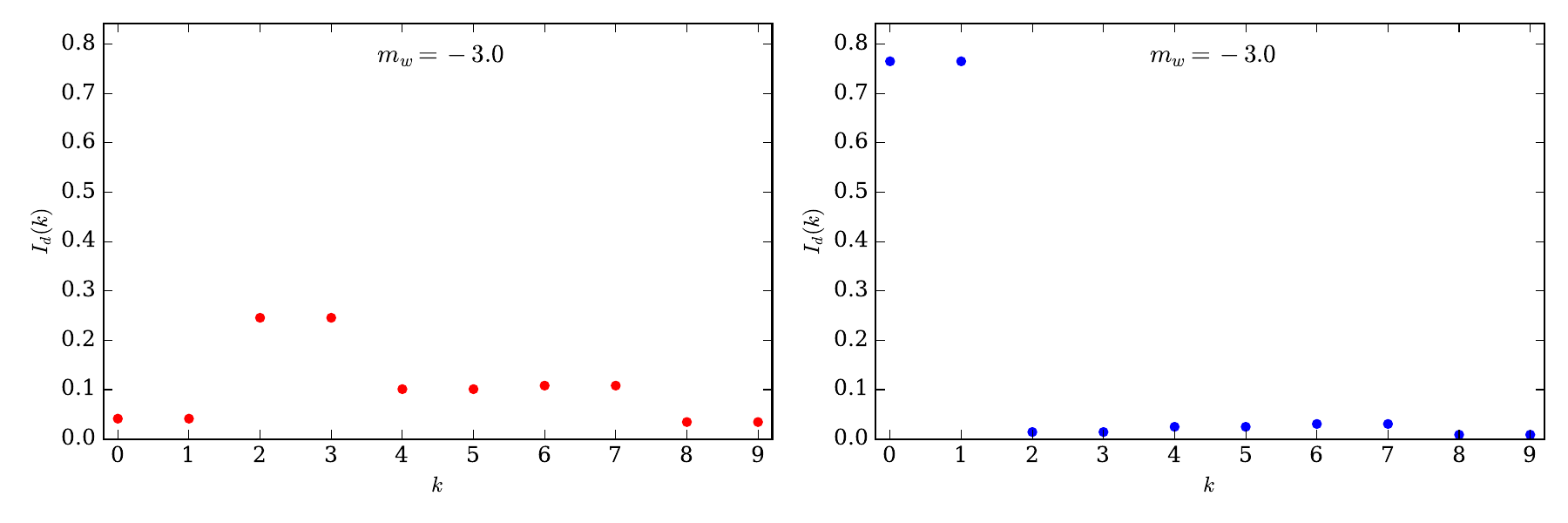}
\includegraphics[scale=0.45]{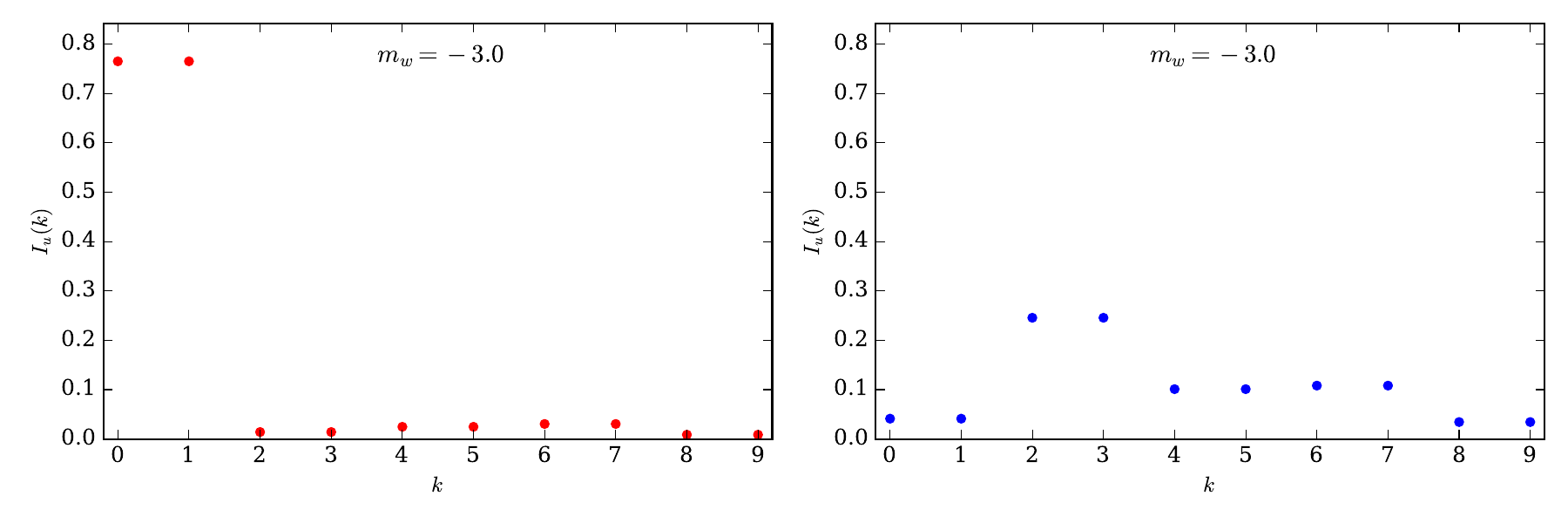}
\caption{Behavior of the low lying spectrum of $W$ as a function of $m_w$ for a fixed monopole-anti-monopole pair.}
\label{fg:wilson-one}
\eef

\section{Results for monopoles}~\label{sec:monopoles}

The monopoles and anti-monopoles are dominated by $\pm 1$ charge in the range of $\beta$ we have studied. As such we focus on $\rho_M(1;L,\beta)$ and we have compiled all the pertinent results in \fgn{action_mono_den_1}. At a fixed gauge coupling, $\beta$, we find a linear dependence of the action density, $\rho_S(L,\beta)$ on $\rho_M(1;L,\beta)$. The data clearly forms a cluster for all values of $L$ considered in this paper at a fixed coupling as seen in the top-left panel of \fgn{action_mono_den_1}. This motivates us to write the empirical relation
\be
\rho_S(L,\beta) = s(\beta) \rho_M(1;L,\beta) + \rho_{S_0}(\beta)\label{monocost}
\ee
where the slope and the intercept in the linear relation do not depend on $L$. 
The slope can be interpreted as the action per monopole-anti-monopole pair since the net monopole charge is always zero. The intercept can be interpreted as the action density at $\beta$ when there are no monopoles. The slope and the intercept are plotted as function of $\beta$ in the top-right panels. These two values were obtained by two different methods. In the first method (pooled), we performed a linear fit of the combined data over all $L$ at a fixed $\beta$. In the second method (averaged), we performed a linear fit of the data at a fixed $L$ and $\beta$ and then averaged the values over different $L$. The results from the two methods are consistent with each other as shown in the figure. There is clear evidence that the slope approaches a finite limit and the intercept does approach zero as $\beta\to\infty$. A linear fit of $s(\beta)$ vs $\frac{1}{\beta}$ was performed in the range highlighted in the bottom two panels and it resulted in $s(\infty)=4.01(3)$.
The clusters in the top-left plot move from the top-right to bottom-left as we increase beta and gets more concentrated at the bottom-left suggesting an exponential dependence of monopole density and action density on $\beta$. To study this quantitatively,
we have plotted 
$\langle \rho_S(L,\beta)\rangle - \rho_{S_0}(\beta)$ and $\langle\rho_M(1;L,\beta)\rangle$ as a function of $\beta$ on the bottom left and right panels. The results at a fixed $L$ and $L\to\infty$ are shown. The $L\to\infty$ limit has been reliably computed in the gray shaded region and both show an exponential dependence with $2c$ in \eqn{massgap} consistent with the linear relation in~\eqn{monocost}.
Along with the estimate for the intercept from the linear fit, we conclude that our data is consistent with
\be
\lim_{L\to\infty}  \langle \rho_M(1;L,\beta)\rangle = 35(6) e^{-4.99(8) \beta}.\label{mdenus}
\ee

We now move on to understand the possibility of attraction and repulsion between monopole charges at short distances using the quantity $g(r)$ defined in \scn{monoobs}. We have plotted $\ln(g(r))$ as a function of $r$ 
in \fgn{ln_g_r}. Three different values of $L$ are shown (largest, third largest and fifth largest $L$). The data for different values of $L$ fall on top of each other for smaller values of $\beta$ and the deviations at $\beta=2.375$ are still quite small. We can use \fgn{ln_g_r} to perform a comparison between the attractive and repulsive strength.   We see that there is a repulsion  between like charge monopoles and an attraction between opposite charge monopoles at short distances. This suggest the formation of dipoles which we will be able to study using the spectrum of Wilson-Dirac operator. Then by looking at $g(r)$ for all monopole pairs regardless of charge, we see that at short distances oppositely charged monopoles dominate suggesting formation of dipoles. Furthermore, we see that charge symmetry is maintained since the effect on positive and negative like-charge monopoles is the same. This is expected since the action is symmetric under charge conjugation $\left(U_\mu({\bf n}) \to U^*_\mu({\bf n})\right)$. The mean distance ($d^\pm$ and $d^{+-}$ defined in \eqn{meandis}) between monopoles (independent of charge) should go like $0.4803 L$ for a dilute random gas and we see that our data is consistent with this result and is independent of $L$. Furthermore, the result in \fgn{ln_g_r} stays the same as $L\to\infty$ and we see that effect of attraction and repulsion changes from $2$ to $8$ as $\beta$ is increased. On that other hand, the mean distance goes to infinity. Therefore, the attraction and repulsion is microscopic in nature and the increase in the typical microscopic distance with $\beta$ where we have attraction and repulsion is just an effect of the increasing separation (decreasing density) of monopoles.

\bef
\centering
\includegraphics[scale=0.35]{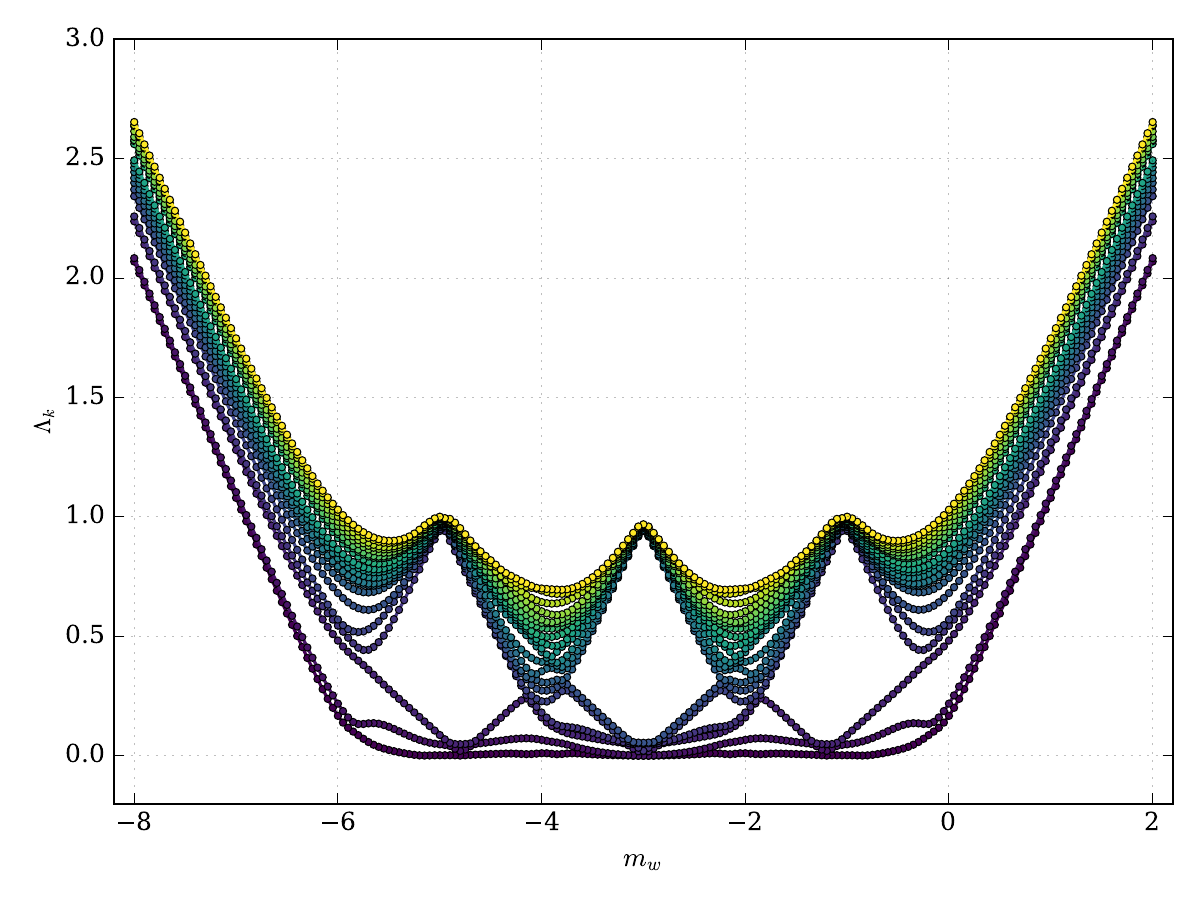}
\includegraphics[scale=0.35]{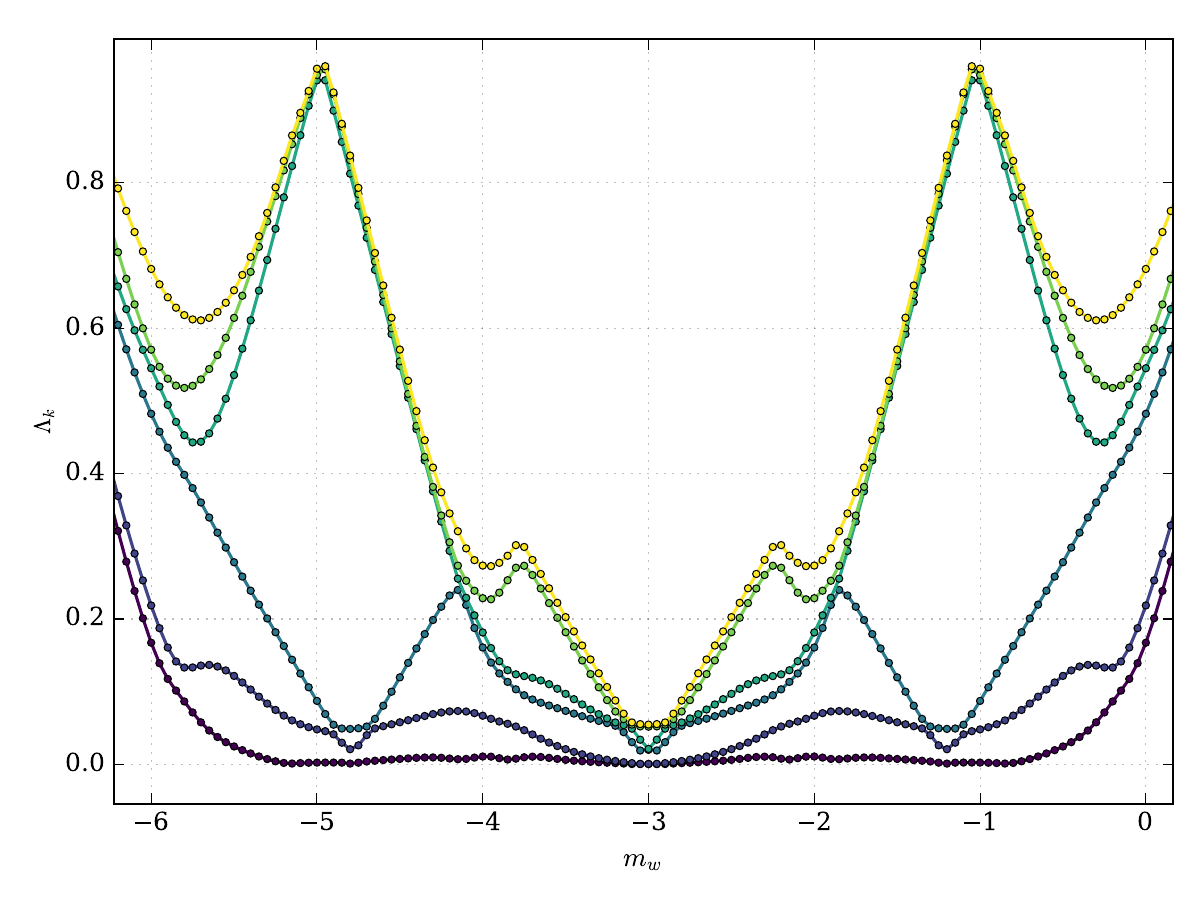}
\includegraphics[scale=0.45]{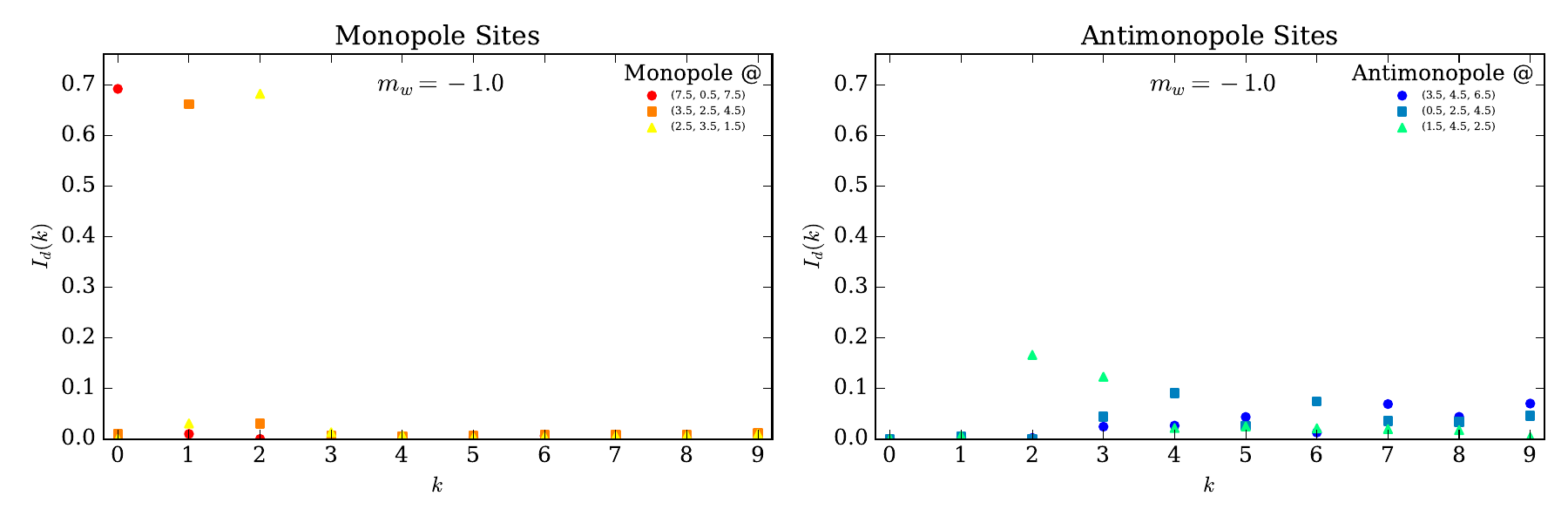}
\includegraphics[scale=0.45]{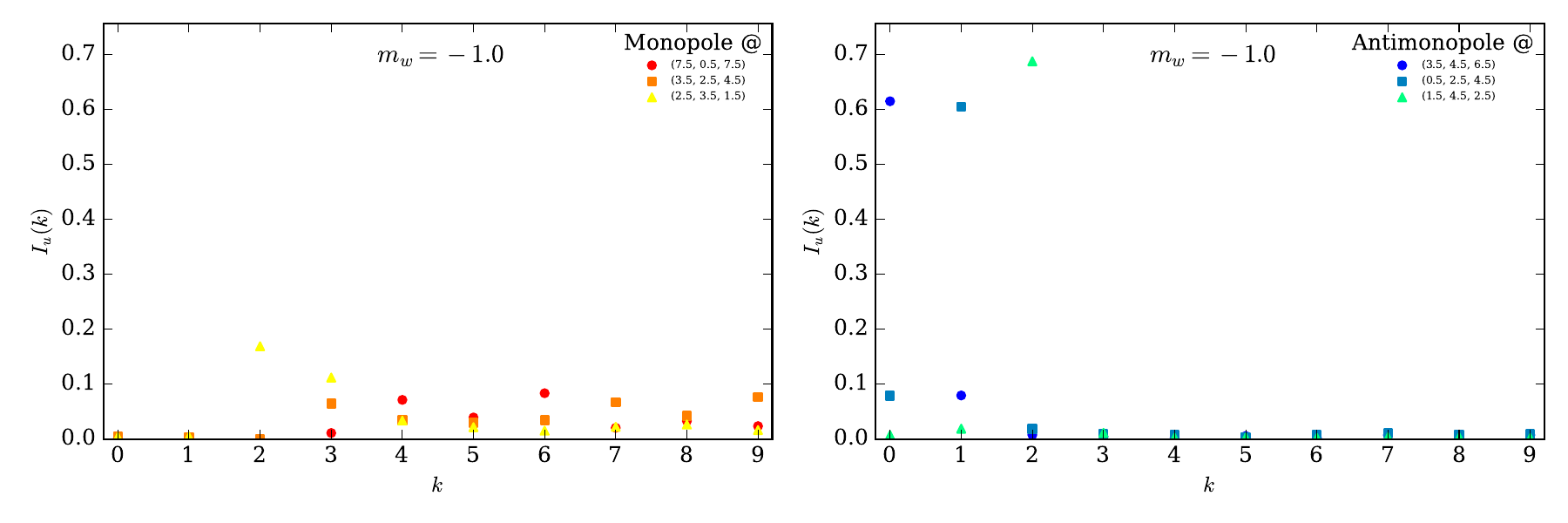}
\includegraphics[scale=0.45]{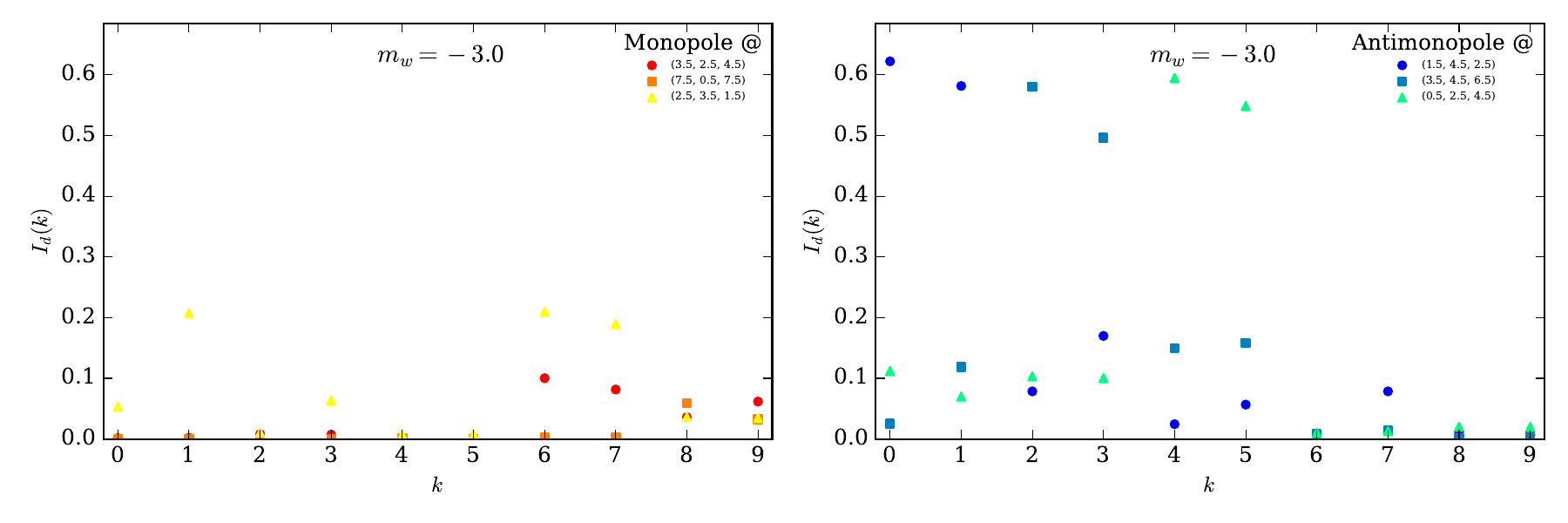}
\includegraphics[scale=0.45]{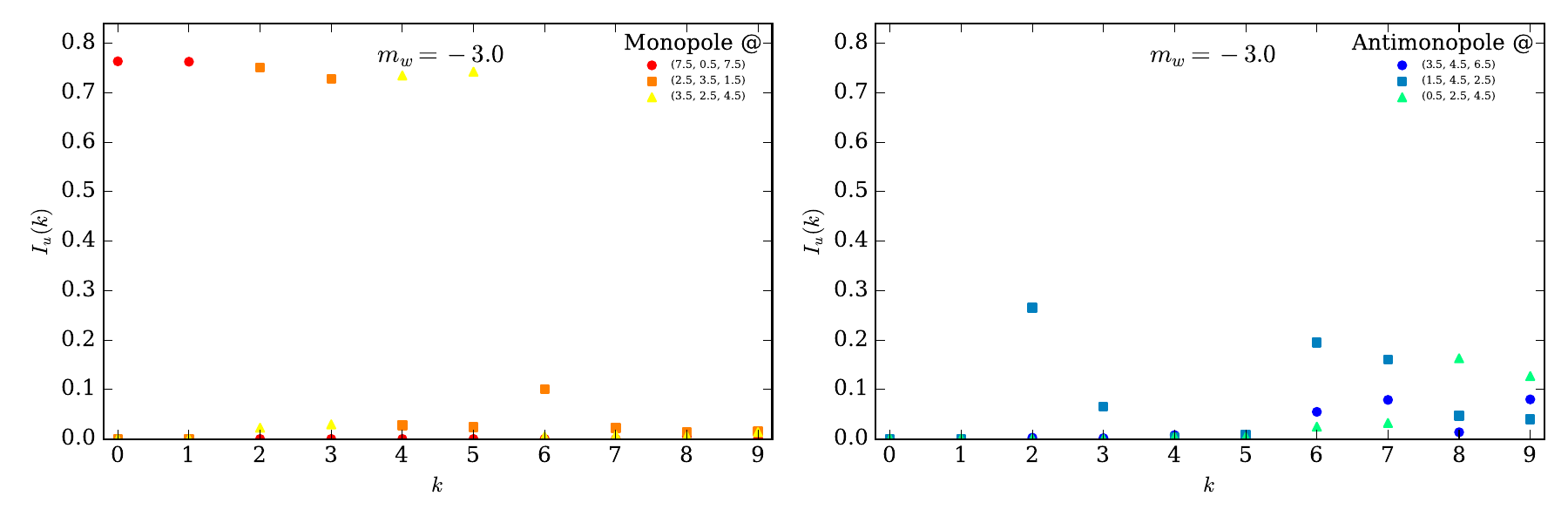}
\caption{Behavior of the low lying spectrum of $W$ as a function of $m_w$ for a thermalized configurations with three monopole-anti-monopole pairs.}
\label{fg:wilson-three}
\eef

\bef
    \centering
    \begin{subfigure}{0.48\textwidth}
        \centering
        \includegraphics[width=\textwidth]{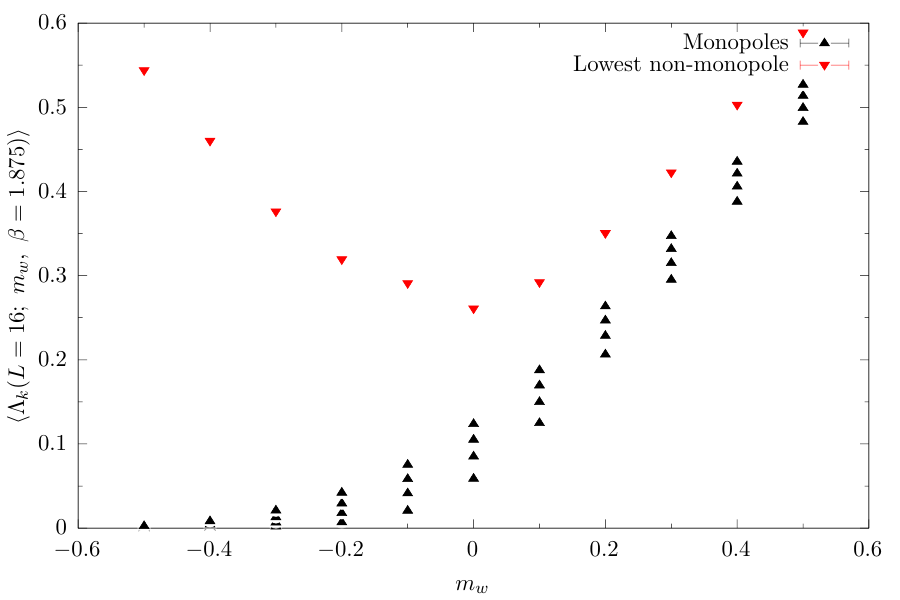}
        \caption{$\beta = 1.875, L =16$}
        \label{fig:lam_mw_2250}
    \end{subfigure}
    \hfill
    \begin{subfigure}{0.48\textwidth}
        \centering
        \includegraphics[width=\textwidth]{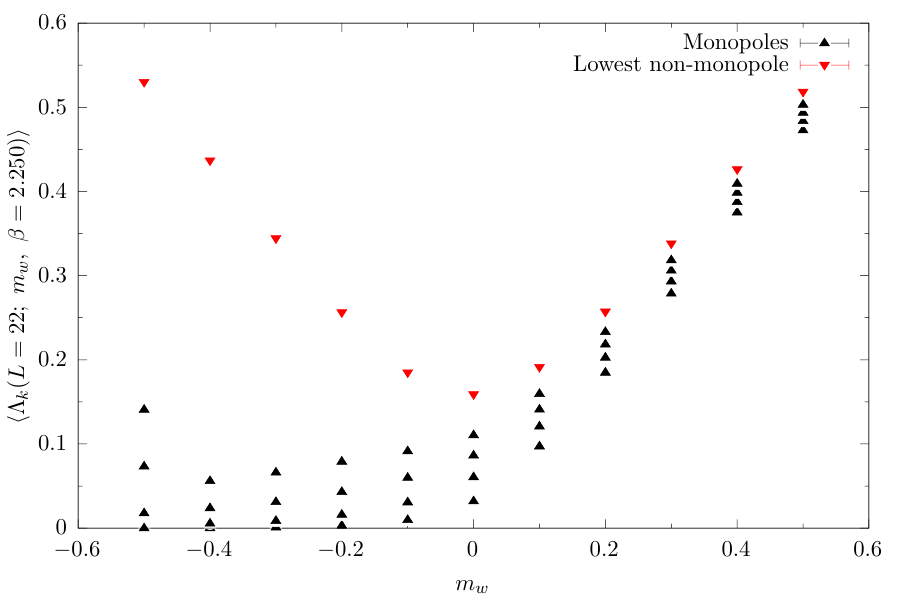}
        \caption{$\beta = 2.250, L = 22$}
        \label{fig:lam_mw_2500}
    \end{subfigure}
    \caption{The eigenvalues $\Lambda_k(L;m_w,\beta)$ as a function of $m_w$ at a fixed value of $L$ and $\beta$.}
    \label{fg:lam_mw}
\eef

\section{Results for the Wilson fermion spectrum}~\label{sec:wilson}

We will develop an understanding of the low lying spectrum of the Wilson-Dirac operator, $W$, defined in \eqn{woper} in this section. We will start with a fixed gauge field background of a well separated monopole-anti-monopole pair as defined in~\cite{Karthik:2019jds}. We will set the Sheikholeslami-Wohlert coefficient, $c_{\rm sw}$, to zero to enable a simple understanding of the spectrum. The pair is separated along the $z$-axis by a distance of $\frac{L}{4}$ and we will present results on an $8^3$ lattice. \fgn{wilson-one} has five rows with a pair of plots in each row and together will help us understand the behavior of the Wilson-Dirac spectrum in a background of a single monopole and anti-monopole. The low lying spectrum (we only show the lowest twenty positive eigenvalues) is shown in the top-left panel of~\fgn{wilson-one} as a function of $m_w \in [-8,2]$. The sole effect of the monopole-anti-monopole pair is the presence of a single anomalously small eigenvalue in the range $m_w\in [-2,0], [-6,-4]$ and two small eigenvalues in the range $m_w\in [-4,-2]$. If the small mode(s) were due to a continuum like configuration, we should find $1,3,3,1$ almost zero modes close to $m_w=0,-2,-4,-6$ respectively. The spectral symmetry around $m_w=-3$ is a property of Wilson fermions in all gauge field backgrounds. The presence of anomalously small eigenvalues away from $m_w=0,-2,-4,-6$ and the break in the degeneracy pattern from a continuum like gauge field suggest that  the effect of the monopole-anti-monopole pair is singular and not continuum-like. The piece-wise linear behavior of the second lowest eigenvalue is due the gauge field background being only made up of a single monopole and anti-monopole and the use of periodic boundary conditions for fermions. The behavior of the two lowest eigenvalues are singled out in the top-right plot and we see, by comparing the scales in the top-left and top-right plot, the presence of a gap separating this eigenvalue from the rest at the masses away from the doubler points. Let us define the intensity of a mode, $k$, at the location of the monopole (or anti-monopole) as the sum of 
\be
I_u(k)=\sum_{{\bf n}\in {\rm cube}} u_k^\dagger({\bf n}) u_k({\bf n});\qquad I_d(k)=\sum_{{\bf n}\in{\rm cube}} d_k^\dagger({\bf n}) d_k({\bf n})
\ee
where the sum is over the sites of the cube that defines the location of the monopole (or anti-monopole). It is assumed that both the vectors, $u_k$ and $d_k$, are normalized to unity. These were computed at $m_w=-1$ and $m_w=-3$. We see clear evidence in the plots of $I_{d,u}(k)$ for $m_w = -1$ of \fgn{wilson-one} that the anomalously small mode is localized at the site of the monopole or anti-monopole when compared to the other modes. Note that the lowest $d$ mode is localized at the monopole and the lowest $u$ mode is localized at the anti-monopole. This breaking of flavor symmetry is due to the non-commutativity of $D_i$ and $D_i^\dagger$ and is specific to Wilson fermions. The localization of small mode and the absence of localization in the mode above the small mode is due to the fact that the high intensity at the location of the monopole (anti-monopole) is only present for the lowest eigenmode. The spectrum at $m_w=-3$ shows two anomalously small eigenvalues that are degenerate as expected. This has two effects on the eigenmodes as seen in $I_{d,u}(k)$ for $m_w = -3$ of~\fgn{wilson-one}. The lowest two modes show a clear peak at the location of the monopole and anti-monopole. Whereas, $I_d$ showed the peak at the location of the monopole at $m_w=-1$, the peaks in $I_d$ are now at the location of the anti-monopole. The reversal of the roles of $u$ and $d$ between $m_w=-1$ and $m_w=-3$ can either be viewed as a flip in the flavor or we can also think of it as {\sl chirality} if we use the four component reducible representation of the Dirac spinors to realize a pair of flavors. In the latter case, we have two other Dirac spin matrices that anti-commute with the three Dirac matrices that appear in the Wilson-Dirac matrices and one of them can be intrepreted as chirality. We could take this heuristic connection between flavor and chirality a bit further if we also use a connection between monopoles in three dimensions and instantons in four dimensions. Instantons result in level crossings~\cite{Edwards:1998gk} in the spectral flow of the Wilson-Dirac operator and we could view the single small mode being analogous to a single instanton level crossing. If so, we would have expected four level crossings of the opposite type in an instanton background close to $m_w=-2$~\cite{Edwards:1998gk}. One of them cancels the previous level crossing resulting in net crossing of three modes. An analogy in three dimensions would then suggest we have two small modes if $-4 < m_w < -2$ and this is consistent with our observation.
Note that we also see a numerical symmetry between the behavior in $I_u$ and $I_d$ when we go from monopole location to anti-monopole location and this is due to the fact that the gauge field background goes to itself under parity.

Now we move on to a similar analysis of a thermalized configuration at $\beta=1.7$ on $8^3$ lattice. This particular configuration has three monopoles and three anti-monopoles located on cubes centered at 
$$\left(\frac{15}{2},\frac{1}{2},\frac{15}{2}\right),\left(\frac{7}{2},\frac{5}{2},\frac{9}{2}\right),\left(\frac{5}{2},\frac{7}{2},\frac{3}{2}\right)  $$
and
$$\left(\frac{7}{2},\frac{9}{2},\frac{13}{2}\right),\left(\frac{1}{2},\frac{5}{2},\frac{9}{2}\right),  \left(\frac{3}{2},\frac{9}{2},\frac{5}{2}\right)$$
respectively. \fgn{wilson-three} has five rows with a pair of plots in each row and together will help us understand the behavior of the Wilson-Dirac spectrum in the background of a thermalized configuration.  The spectrum shows three anomalously small eigenvalues (on the positive side) for $m_w\in [-2,0],[-6,-4]$ and six small eigenvalues for $m_w\in [-4,-2]$. The top-right panel zooms in on the six lowest eigenvalues seen in the top-left panel of \fgn{wilson-three} and we clearly see the anomalously small eigenvalues in $m_w\in [-6,0]$. Note that at $m_w=-3$, we observe three degenerate pairs of anomalously small eigenvalues. The intensities at the location of the monopole and anti-monopole sites are shown for $I_{d,u}$ at $m_w=-1,-3$. Again, $I_d$ peaks at monopoles and $I_u$ peaks at anti-monopoles for $m_w =-1$ showing a breaking of flavor symmetry. Furthermore the roles of $u$ and $d$ are reversed for $m_w = -3$ when compared with $m_w=-1$. In addition, the location of the observed peaks matches the pair-wise degeneracy at $m_w=-3$. Note that we have two more features in the $m_w=-1$ and $m_w=-3$ pair of plots. Due to the localized nature of the eigenmodes associated with the anomalously small eigenvalues, each of these modes show a peak in $I_{d,u}$ at a particular monopole and anti-monopole in the order of the location listed above. Since the eigenvalues of $D_iD_i^\dagger$ and $D_i^\dagger D_i$ are paired (c.f.~\eqn{eigw}), we see that these modes naturally pair a monopole with an anti-monopole in the order listed above.  These observations are based on a single sample configuration but they are intriguing and suggests that the anomalously small eigenmodes of the Wilson-Dirac operator identifies the number of monopoles; their locations and charges; and   the pairing of monopoles and anti-monopoles.

We now analyze the low lying spectrum at different values of $L$ and $\beta$. For this purpose we use a tuned Sheikholeslami-Wohlert parameter, $c_{\rm sw}$.  We will compute the spectrum for several values of $m_w \in [-0.5,0.5]$ on each gauge field background and therefore the results at different values of $m_w$ are fully correlated. To start, we plot  the lowest four anomalously small eigenvalues that arise from monopoles along with the lowest  eigenvalue  that is separated from the small eigenvalues by a gap for $m_w < 0$ as a function of the Wilson mass in \fgn{lam_mw}. The anomalously small eigenvalues can be seen to get smaller as $m_w$ becomes more negative whereas the  eigenvalue that sets the gap behave quadratic in $m_w$ as expected. We tuned $c_{\rm sw}$ so that the minimum of the quadratic behavior occurred close to $m_w=0$.  The two panels in \fgn{lam_mw} show the behavior at two different couplings and we note that there are many more small eigenvalues at $\beta=1.875$ than have been shown. 

\bef 
\includegraphics[scale=0.5]{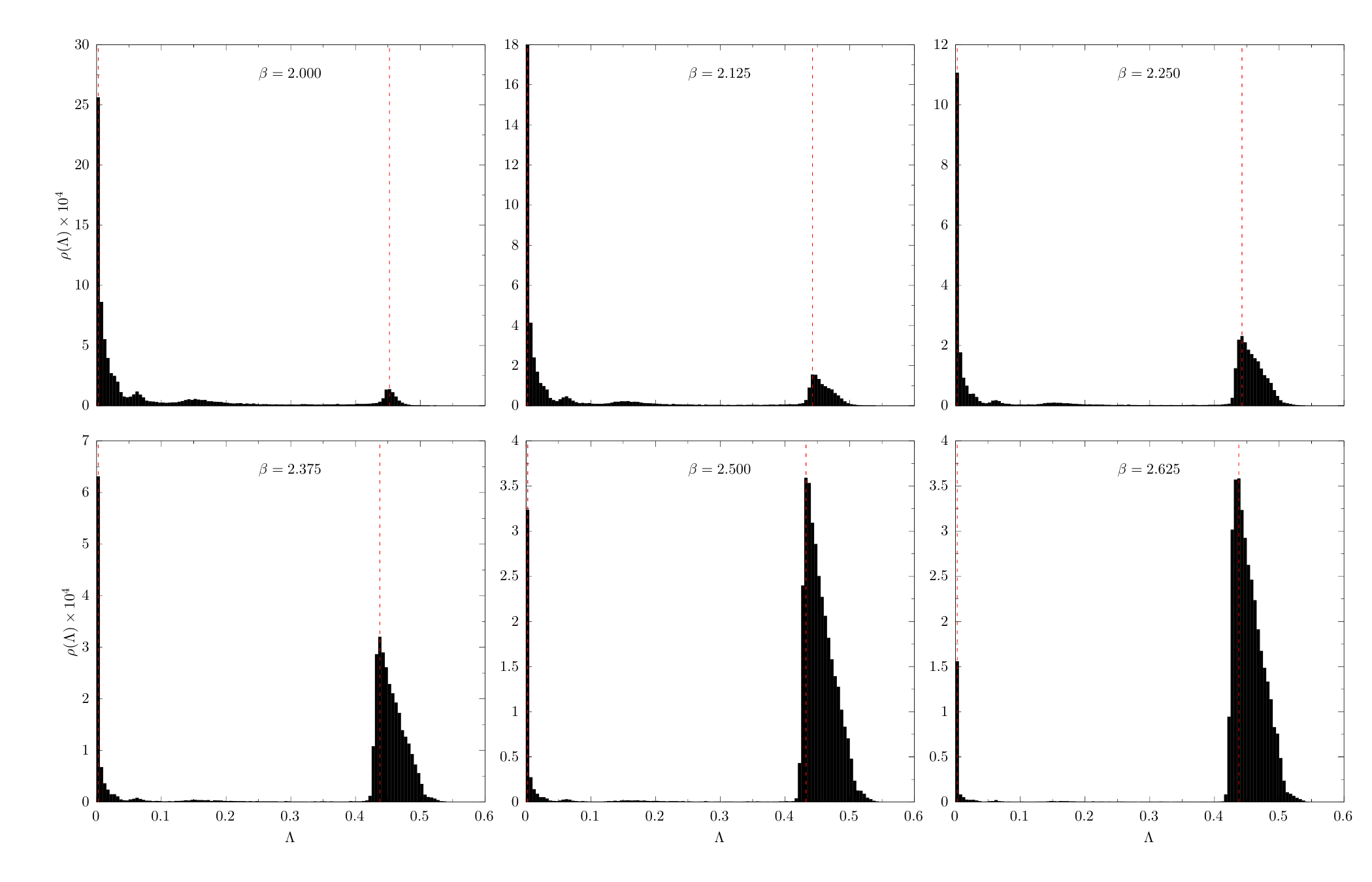}
\caption{$\rho(\Lambda)$ for $m_w=-0.4$ and various $\beta$. Peaks are distinguished by dashed lines.} 
\label{fg:eig-dist}
\eef

The numerical computation of the spectrum only found a fraction of the low lying eigenvalues to reduce the computational time. The fraction that was computed depended on the expected number of small modes from the number of monopoles as measured by~\eqn{gmono}. The spectral distribution, $\rho(\Lambda)$, is shown for six different lattice gauge couplings in \fgn{eig-dist} with the normalization set to unity if the entire spectrum had been computed. The cutoff at the high end is due to the fact we did not compute the entire spectrum. The distribution includes the spectrum computed at all values of $L$ and we have fixed the wilson mass to $m_w=-0.4$.  There is a clear gap that separates the anomalously small eigenvalues from the rest. Due to the presence of many more anomalously small eigenvalues at smaller $\beta$, the gap is not as well defined but the peak at $\Lambda=0$ is larger for the same reason. We compared the number of anomalously small eigenvalues with the number of monopoles computed using~\eqn{gmono}. Whereas there is an overwhelming agreement between the two, we also found some disagreements. Smeared gauge fields were used in~\eqn{gmono} and in the Wilson-Dirac operator. The disagreements usually arose in situations where a monopole and anti-monopole were separated by a single lattice spacing. Had we defined~\eqn{gmono} using $2\times 2$ plaquettes on a $2^3$ cube, we would have found a small difference in the number of monopoles. The spectrum of Wilson-Dirac operator depends on the entire gauge field configuration on the lattice. This difference is the cause for the minor disagreements between the two methods for defining the number of monopoles. Since we seem to be able to identify the location of the monopoles and anti-monopoles by looking at intensity peaks in the eigenmodes we may use these modes as a method to identify all monopoles and anti-monopoles including their locations.

\bef 
\includegraphics[scale=0.5]{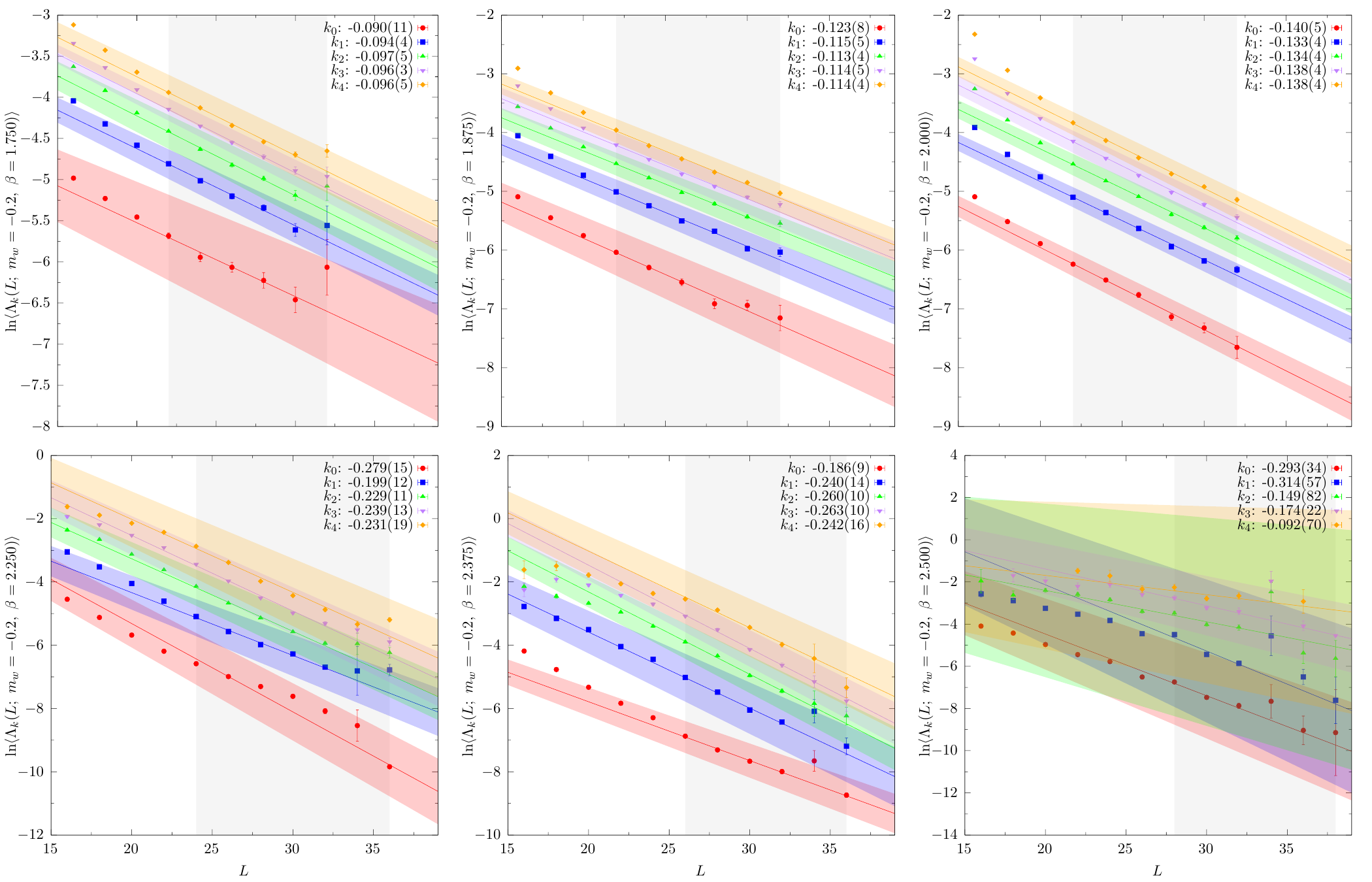}
\caption{$\ln\Lambda_K(L;m_w, \beta)$ as a function of $L$ for six different values of $\beta$ at $m_w=-0.2$.} 
\label{fg:ln_lam_neg_mw}
\eef

The behavior of the small eigenvalue associated with the monopole in a fixed gauge field background~\cite{Karthik:2019jds} was shown to fall off exponentially with $L$ for $m_w < 0$ and the rate of fall was shown to reduce as $m_w$ increased toward zero. We therefore assume that
\be
\langle \Lambda_k(L;m_w,\beta)\rangle = c_{k,\Lambda}(m_w,\beta) e^{-M_\Lambda(m_w,\beta) L}\label{eigvsL}
\ee
and we have assumed that the rate of fall does not depend on the specific mode.
We show the behavior of the five lowest eigenvalues as a function of $L$ for a range of $\beta$ at $m_w=-0.2$ in \fgn{ln_lam_neg_mw}. Indeed, we see that the rate of fall does not depend on the specific mode. It should be pointed out that finite $L$ effects beyond the exponential fall-off are present at values of $L$ considered at the largest $\beta$ and this is consistent with the need to go to larger $L$ at larger $\beta$ to see asymptotic behavior. Next, we point out an effect that might not seem obvious. The rate of fall increases with $\beta$. The gap is clearly larger at smaller $\beta$ as shown in \fgn{lam_mw}. But we also have more monopoles at smaller $\beta$ and we need to go to larger $L$ at higher $\beta$ to see the asymptotic behavior. 
The left panel of \fgn{ln_lam_neg_mw_fn_mw_rn_beta} shows two effects for $M_\Lambda(m_w,\beta)$ at $\beta=1.875$. First of all, we confirm the assumption that it does not depend on the mode. We also see that the behavior as a function of $m_w$ at a fixed $\beta$ is consistent with
\be
M_\Lambda(m_w,\beta) = s_\Lambda(\beta) m_w + b_\Lambda(\beta),\label{massvsmw}
\ee
and this matches the behavior found in a fixed monopole-anti-monopole background in~\cite{Karthik:2019jds}. The right-panel shows the behavior of $M_\Lambda(m_w,\beta)$ of the lowest mode as a function of $m_w$ for a range of $\beta$. From this we conclude that $s_\Lambda(\beta)$ increases with $\beta$ in the range where asymptotic scaling in $L$ is observed.

\bef
    \includegraphics[scale=0.5]{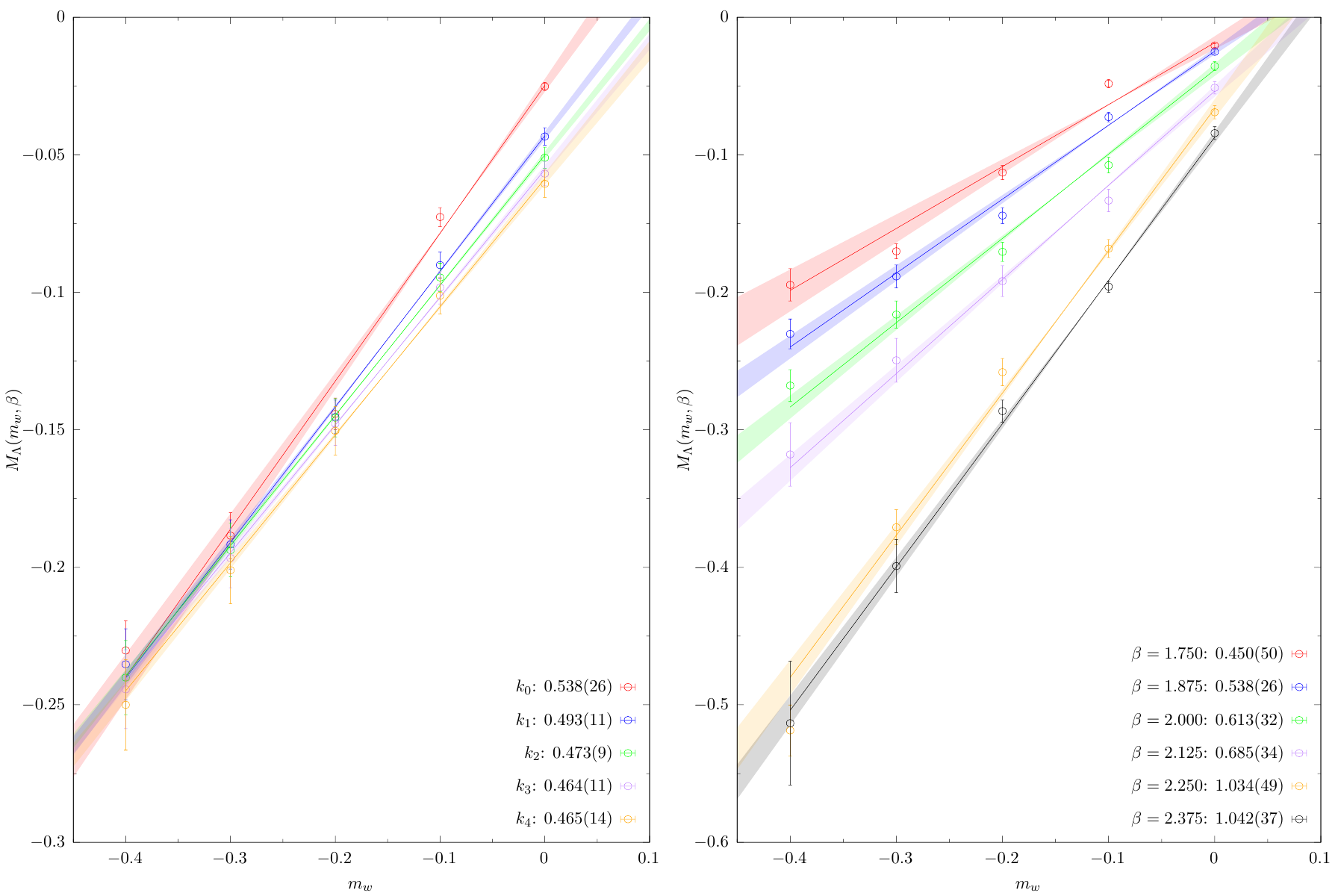}
    \label{ln_lam_neg_mw_fn_mw_rn_id}
    \caption{ $M_\Lambda(m_w,\beta)$ as a function of $m_w$. The left panel shows that this quantity does not depend on the mode itself. The right panel shows that linear behavior in \eqn{massvsmw} and the increasing slope with $\beta$.}
    \label{fg:ln_lam_neg_mw_fn_mw_rn_beta}
\eef

Based on \eqn{eigvsL} and \eqn{massvsmw}, we can define the massless point as the value of $m_w$ where $M_\Lambda(m_w,\beta)$ goes to zero.
The plot in the right panel of \fgn{ln_lam_neg_mw_fn_mw_rn_beta} show that $b_\Lambda(\beta)$ is not zero but its effect is small enough that we could set the massless point to be $m_w=0$. We, therefore, ask how $\Lambda(L;0,\beta)$ behaves as a function of $L$ by setting $M_\Lambda(0,\beta)=0$ and the results are plotted in \fgn{lnlam_vs_lnL}. If there is a fermion bilinear condensate of the conventional type, we expect the eigenvalues to behave as $\frac{1}{L^3}$ and perturbations from free behavior would suggest a behavior of the form $\frac{1}{L}$. The two sets of dashed lines in each of the plot for each mode compares the data to a $\frac{1}{L}$ and $\frac{1}{L^3}$ behavior. There is no evidence for a $\frac{1}{L^3}$ at the smaller values of $\beta$. In fact the data at $\beta=1.750, 1.875, 2.000$ and $2.125$ favor a $\frac{1}{L}$ behavior. There might be some evidence for an emerging $\frac{1}{L^3}$ behavior at $\beta=2.250$ and $\beta=2.375$ at larger $L$ for the third and fourth modes. One possible explanation is that monopole based modes do not support a $\frac{1}{L^3}$ behavior where as the bulk (non-monopole) modes support a $\frac{1}{L^3}$ behavior as $\beta\to\infty$. A careful analysis to disentangle these two possibly contrasting behavior is beyond the numerical capability of this paper. It is worth pointing out that a single pair of well separated monopole and anti-monopole results in one small eigenvalue that behaves as $\frac{1}{L}$~\cite{Karthik:2019jds} whereas fermions in the background of gauge fields generated by a non-compact gauge action results in a condensate due to a $\frac{1}{L^3}$ behavior~\cite{Karthik:2017hol}.

\bef
    \includegraphics[width=\textwidth]{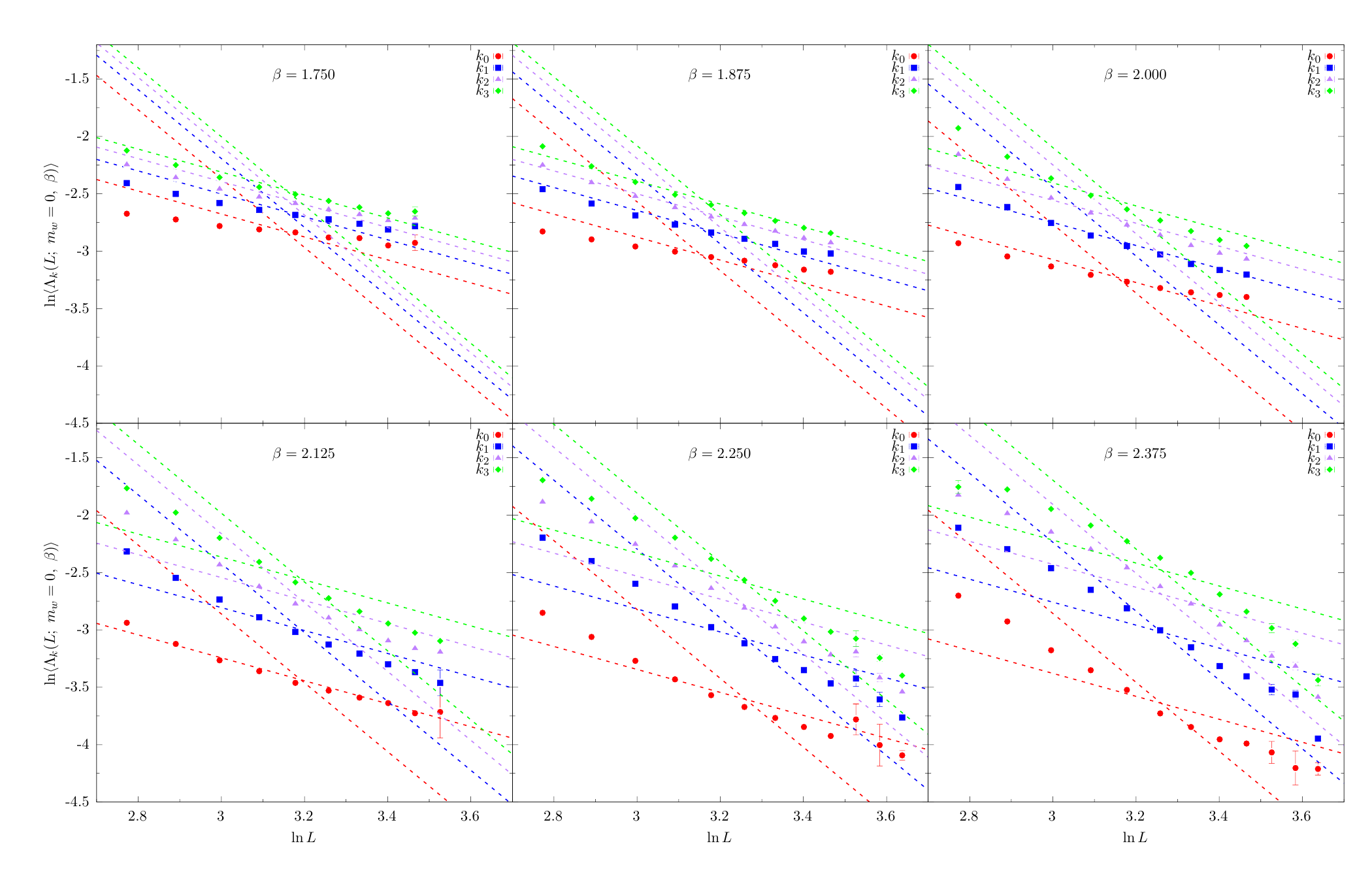}
    \label{lnlam_vs_lnL}
    \caption{ $\ln \langle\Lambda_k(L;0,\beta)\rangle$ as a function of $\ln L$. The two dashed lines are for $\frac{1}{L}$ and $\frac{1}{L^3}$ behavior.}
    \label{fg:lnlam_vs_lnL}
\eef

\section{Conclusions}
We have studied the physics of monopoles in three dimensional compact $U(1)$ gauge theory by using Wilson fermions as a probe. It has been clearly argued in~\cite{Athenodorou:2018sab} that one cannot couple gauge fields to fermions if one has a finite density of monopoles in the continuum limit. If the only relevant continuum theory is where we set the lattice gauge coupling to $\beta = \frac{2L}{\ell}$ on an $L^3$ lattice and take the continuum limit as $L\to\infty$ at a fixed $\ell$ which corresponds to the dimensionless size of the periodic box measured in units of the continuum gauge coupling, then monopoles are absent and one can study QED in three dimensions with a non-compact gauge action. This has been done using lattice regularization~\cite{Karthik:2015sgq,Karthik:2016ppr,Karthik:2017hol} to show that parity invariant theories are scale invariant and also to compute the scaling dimensions of monopole operators~\cite{Karthik:2019mrr,Karthik:2024ffr}. In this context, it could be argued that the study performed here is not of physical relevance if the aim is to eventually couple the compact $U(1)$ gauge theory to fermions. Yet, our analysis does reveal an interesting interplay between monopoles and Wilson-Dirac fermions that has a potential effect in a study using overlap fermions.

A single two-component fermion flavor has a parity anomaly and the Wilson-Dirac operator, $D_i$, realizes this properly~\cite{Coste:1989wf}. This becomes even more evident if we note that
\be
V=D_i \frac{1}{\sqrt{D_i^\dagger D_i}};\qquad VV^\dagger = V^\dagger V = \mathbf{I}\label{voper}
\ee
is the unitary operator used to define a single two-component fermion operator, $D_o=\frac{1+V}{2}$.
The fermion determinant of a parity invariant two flavor theory is
\be
\det \frac{1+V}{2} \det \frac{1+V^\dagger}{2} = {\det}^2\frac{1+V}{2} \det V^\dagger
\ee
and we see that a Chern-Simons term, $\det V^\dagger$ cancels the parity anomaly of two flavors.
The $O(2)$ flavor symmetry present in a two-flavor parity invariant theory realized on the lattice using overlap fermions is absent if we use 
\be
\det D_i \det D_i^\dagger
\ee
instead. The main difference is the non-commutativity of $D_i$ with $D_i^\dagger$ on the lattice which is expected to be absent in the continuum limit. Monopoles are fundamentally singular and their presence does not support this expectation and we find that flavor symmetry is broken if we realize a two-flavor theory using \eqn{woper}.

After establishing the expected asymptotic scaling of the  density of monopoles in \eqn{mdenus}, we show that the collection of monopoles and anti-monopoles behaves like a dilute random gas at distances that define the average distance between the monopole (positive charge) and anti-monopole (negative charge), namely, $d^{\pm}=d^{+-}=0.4803L$ independent of $\beta$. We do observe a repulsion between like charged objects and an attraction between unlike charged objects at microscopic distances (these remain fixed at a fixed $\beta$ as $L\to\infty$) and the microscopic distances reflect the decreasing density of monopoles with increasing $\beta$. By performing a configuration-by-configuration analysis of the action density versus the monopole density, we are able to conclude that the action per monopole-anti-monopole pair is $4.01(3)$ in the $\beta\to\infty$ limit.

The primary aim of the paper is to establish parity breaking by monopoles in the spectrum of a pair of spectator Wilson fermions defined in a way to not have a parity anomaly. To this end, we start by studying the low lying spectrum(c.f.~\eqn{eigw}) in a background of a well separated monopole and anti-monopole that was used as a background in~\cite{Karthik:2019mrr,Karthik:2024ffr} to extract the scaling dimension of a monopole in an otherwise non-compact gauge field background. We find one anomalously small eigenvalue, $\Lambda_0 > 0$, for the Wilson mass in the range $m_w\in [-2,0],[-6,-4]$ and two small eigenvalues in $m_w\in [-4,-2]$. This implies that anomalously small modes exist for all values of one sign of the fermion mass noting that the pair of Wilson fermions has $\pm m_w$ as their mass parameter. The  eigenmode components $u_0$ and $d_0$ associated with the two flavors couples to the anti-monopole and monopole respectively in a strong manner showing that parity is broken. 
The doubling of the modes in $m_w\in [-4,-2]$ is seen in the peaks of the eigenmodes and in addition there is reversal in the roles played by the two flavors when we compare the behavior at $m_w=-1$ and $m_w=-3$. It would be numerically time consuming to show the same in every thermalized configuration at a fixed $\beta$ and $L$ but we show this to be the case in a single sample configuration. Moreover, the number of anomalously small modes in $m_w\in [-2,0]$ match the number obtained from \eqn{gmono} and a particular mode is coupled to a particular monopole in such a way that the pair of $u_k$ and $d_k$ associated with the same eigenvalue matches a monopole with an anti-monopole that minimizes the sum of the distances between the pairs. 
But we do see discrepancies between the number of anomalously small eigenvalues and the number of monopoles as defined by \eqn{gmono} when a monopole is only a single lattice spacing away from an anti-monopole. We attribute this to the fact that the presence of anomalously small eigenvalues takes into account the full gauge field configuration on the lattice.

By studying the spectrum of the Wilson-Dirac operator over a wide range of $\beta$ and $L$ we establish that the anomalously small eigenvalues exponentially go to zero with $L$ at a fixed $\beta$ and there is a clear gap between this part of the spectrum and the bulk of the spectrum not attributed to the monopoles. If we were to define a fermion bilinear condensate using
\be
\Sigma_f(\beta) = \lim_{L\to\infty} \frac{1}{L^3}\Tr \frac{1}{W},
\ee
we will find an infinite valued condensate for $-6 < m_w < 0$. Since the gap between the anomalously small eigenvalues and the rest vanishes as $m_w\to 0$, it is possible a finite valued condensate is realized as $m_w\to 0$. Our attempt to show this did not fare well due to the fact that the low lying eigenvalues did not show a $\frac{1}{L^3}$ dependence at the values of $L$ and $\beta$ we studied. It is possible that eigenmodes slightly above the low lying eigenvalues show a $\frac{1}{L^3}$ behavior but one needs to go to larger $\beta$ and $L$ to establish this. This is beyond the numerical scope of this paper and we conclude that the presence of a finite density of monopoles, however small, makes it difficult to establish the presence of a finite condensate that was shown to be present in a non-compact abelian gauge theory~\cite{Karthik:2017hol} where monopoles are absent.

It would be interesting to repeat the analysis for the eigenmodes using overlap fermions since it preserves the parity symmetry on the lattice. But it will be technically difficult since the operator, $V$, in \eqn{voper} has the inverse of $D_i^\dagger D_i$. A careful study of the low-lying eigenmodes in a full ensemble will be interesting since it will enable us to understand how a finite density of monopoles affects the structure of the eigenmodes. If one wishes to study a physical theory with monopoles it would be better to consider a $SU(2)$ gauge theory with a scalar field in addition to a pair of fermions. 
\acknowledgments
 R.N. acknowledges partial support by the NSF under grant number
PHY-2310479.  This work used Expanse at SDSC through allocation
PHY240084 and PHY260214 from the Advanced Cyberinfrastructure Coordination
Ecosystem: Services \& Support (ACCESS) program, which is supported
by National Science Foundation grants \#2138259, \#2138286, \#2138307,
\#2137603, and \#2138296.

\bibliography{biblio}
\end{document}